\documentstyle[equations,12pt,epsf]{article}
\begin{document}

\begin{flushright}
IPNO/TH 96-32
\end{flushright}
\vspace{1 cm}
\begin{center}
{\large {\bf Gauge transformations in relativistic two-particle 
\protect \\
constraint theory}}
\vspace{1.5 cm}

H. Jallouli\footnote{e-mail: jallouli@ipncls.in2p3.fr}
and H. Sazdjian\footnote{e-mail: sazdjian@ipno.in2p3.fr}\\
\renewcommand{\thefootnote}{\fnsymbol{footnote}}
{\it Division de Physique Th\'eorique\footnote{Unit\'e de Recherche des
Universit\'es Paris 11 et Paris 6 associ\'ee au CNRS.},
Institut de Physique Nucl\'eaire,\\
Universit\'e Paris XI,\\
F-91406 Orsay Cedex, France}\\
\end{center}
\newpage

\begin{center}
{\large Abstract}
\end{center}

Using connection with quantum field theory, the infinitesimal covariant 
abelian gauge transformation laws of relativistic two-particle
constraint theory wave functions and potentials are established 
and weak invariance of the corresponding wave equations shown.
Because of the three-dimensional projection operation, these transformation
laws are interaction dependent. Simplifications occur for local
potentials, which result, in each formal order of 
perturbation theory, from the infra-red leading effects of multiphoton 
exchange diagrams. In this case, the finite gauge transformation can
explicitly be represented, with a suitable approximation and up to a
multiplicative factor, by a momentum dependent unitary operator that
acts in $x$-space as a local dilatation operator.
The latter is utilized to reconstruct from the Feynman gauge the
potentials in other linear covariant gauges. The resulting effective 
potential of the final Pauli-Schr\"odinger type eigenvalue equation has 
the gauge invariant attractive singularity $\alpha^2/r^2$, leading to a 
gauge invariant critical coupling constant $\alpha_c =1/2$.
\par
PACS numbers: 03.65.Pm, 11.10.St, 12.20.Ds, 11.80.Fv.
\par
Keywords: Relativistic wave equations, Bethe-Salpeter equation,
constraint theory, gauge transformations.
\par

\newpage

\section{Introduction}

The knowledge of the behavior of Green's functions under gauge
transformations in QED \cite{lk,zjz,o} plays a crucial role in proving
gauge invariance of observables and in particular of bound state 
energies \cite{by,ffh}. In practical calculations, however, one generally 
uses approximations to exact equations, which necessitate a close control
of the degree of realization of the various symmetries of the system 
under study. 
\par
In QED, it has appeared that the Coulomb gauge is the most convenient
gauge for treating the bound state problem, since it allows the optimal
expansion of the Bethe-Salpeter equation \cite{sb,gml,nnns} around the
nonrelativistic theory \cite{by,n,lcl,brbr,m}. Covariant gauges produce,
at each formal order of perturbation theory, spurious infra-red 
singularities that are cancelled only by higher order diagrams 
\cite{bcm,l} and thus become of less practical interest. The main 
disadvantage of the Coulomb gauge is, however, its noncovariant nature,
which does not allow its incorporation in covariant equations.
\par
From this viewpoint, constraint theory, which leads to a manifestly
covariant three-dimensional description of two-body systems 
\cite{ct,cva1,s}, has opened a new perspective in the subject. It was
shown \cite{t2,rta,js} that the expansion of the Bethe-Salpeter equation
around the constraint theory wave equations in the Feynman gauge (as well
as for scalar interactions) is free of the abovementioned diseases of
covariant gauges and allows a systematic study of infra-red leading effects
of multiphoton exchange diagrams; the latter can then be represented in 
three-dimensional $x$-space as local potentials. Summing the series of 
these leading terms one obtains a local potential in compact form 
\cite{js}, which is well suited for a continuation to the strong coupling
domain of the theory or for a generalization to other effective 
interactions.
\par
The purpose of the present paper is to investigate the forms of the
local potentials in linear covariant gauges and to find the
relationships between them. Our approach is accomplished in two steps.
First, using the connection of the constraint theory wave equations
with the Bethe-Salpeter equation, we determine the infinitesimal gauge
transformation laws of constraint theory wave functions and potentials
for fermion-antifermion systems and establish the (weak) invariance of
the corresponding wave equations. Second, specializing to the local
potential approximation, we show that the finite gauge transformation
can be represented, with a suitable approximation and up to a 
multiplicative factor, by a momentum dependent unitary operator that
acts in $x$-space as a local dilatation operator. 
Under the action of the latter, the potentials undergo two kinds of
modification. In the first one, they are changed in a form invariant
way, by the
replacement of their argument $r$ (the c.m. interparticle distance) 
by a function $r(\xi)$, where $\xi$ is the gauge variation parameter.
Apart from a rapid variation near the origin, $r(\xi)$ is essentially
dominated by its large-distance behavior, in which $r$ is simply
shifted by a constant value. In the second kind of modification,
certain parts of the spacelike components of the electromagnetic
potential are functionally modified.
The combination of these two types of modification allows, in particular,
the reconstruction of the potentials in linear covariant gauges from the
knowledge of the potential in the Feynman gauge.
\par
Studying the short-distance behavior of these (three-dimensional) 
potentials we find that the effective potential that appears in the final 
Pauli-Schr\"odinger type eigenvalue equation has the gauge-invariant
attractive singularity $\alpha^2/r^2$. As is known, to such a singularity
there corresponds a critical value $\alpha_c$ of the coupling constant, 
$\alpha_c =1/2$, at which value the fall to the center phenomenon occurs.
\par
Constraint theory thus provides us with a three-dimensional framework in
which the local potential approximation consistently fulfils the requirement
of gauge invariance of the theory. These results, while deduced in the 
framework of QED, might also survive, with appropriate adaptations, to the
incorporation or consideration of other types of interaction.
\par
The plan of the paper is the following. In Sec. 2, the gauge transformation
properties of the Green's functions and of the Bethe-Salpeter wave function
are reviewed. In Sec. 3, the infinitesimal gauge transformation laws of the
constraint theory wave functions and potentials are determined. The case
of the local approximation of potentials is considered in Sec. 4. The
general properties of the gauge transformations in the local approximation
are studied in Sec. 5. A summary and concluding remarks follow in Sec. 6.
\par

\newpage

\section{Gauge transformations of Green's functions and wave functions}
\setcounter{equation}{0}

Gauge transformation laws of Green's functions in QED can be obtained with 
several equivalent methods: i) by considering the operator changes in the 
field operators \cite{lk}; ii) by modifying in the functional integral the 
gauge fixing condition \cite{zjz}; iii) by using the Ward-Takahashi identities
\cite{o}. In the present work we shall consider only linear covariant 
gauges, characterized by a parameter $\xi$; the photon propagator is then:
\begin{equation} \label{2e1}
D_{\mu\nu}(k)\ =\ -(g_{\mu\nu}-\xi \frac {k_{\mu}k_{\nu}}{k^2})
\frac {i}{k^2+i\epsilon}\ .
\end{equation}
The reference gauge is taken here to be the Feynman gauge and the 
transformations express a Green's function calculated in the gauge $\xi$
with respect to its value in the Feynman gauge ($\xi =0$). More
generally, these transformations concern the passage from a gauge $\xi_1$
to a gauge $\xi_2$, with $\xi=\xi_2-\xi_1$.
\par
The transformation law of the unrenormalized Green's function of a charged
particle (a boson or a fermion) is:
\begin{equation} \label{2e2}
G_{\xi}(x)\ =\ \exp\{ ie^2\xi (\Delta(x)-\Delta(0))\}\ G(x)\ ,
\end{equation}
where $e$ is the unrenormalized electric charge of the particle and
$\Delta$ is defined as:
\begin{equation} \label{2e3}
\Delta (x)\ =\ \int \frac{d^4k}{(2\pi)^4} \frac{e^{{\displaystyle ik.x}}}
{(k^2+i\epsilon)^2}\ ,\ \ \ \ \ \Delta (k)\ =\ \frac{1}{(k^2+i\epsilon)^2}
\ .
\end{equation}
Notice that the transformation law is spin independendent.
\par
For a two-particle Green's function, with particle 1 representing a
fermion (boson) with charge $e_1$ and particle 2 an antifermion (boson)
with charge $-e_2$, the transformation law is \cite{o,bcm}:
\begin{eqnarray} \label{2e4}
G_{\xi}(x_1,x_2;y_1,y_2) &=& \exp\{ i\xi [e_1^2 (\Delta
(x_1-y_1)-\Delta(0)) + e_2^2 (\Delta(x_2-y_2))-\Delta(0))]\}\nonumber \\
& & \times \exp\{ i\xi e_1e_2 [\Delta(x_1-x_2)+\Delta(y_1-y_2)
-\Delta(x_1-y_2)-\Delta(y_1-x_2)]\}\nonumber \\
& & \times G(x_1,x_2;y_1,y_2)\ .
\end{eqnarray}
\par
One also deduces from Eq. (\ref{2e4}) the gauge transformation law of
the Bethe-Salpeter wave function. We assume that the two-particle
system is neutral in charge, hence:
\begin{equation} \label{2e5}
e_1\ =\ e_2\ =\ e\ .
\end{equation}
The time separation between outgoing and ingoing states is defined as
$T = ((x_1^0+x_2^0)-(y_1^0+y_2^0))/2$. 
For large time separations ($T\rightarrow \infty$) and finite values of 
$x_1^0-x_2^0$ and $y_1^0-y_2^0$, $G$ satisfies the cluster decomposition
\cite{gml}:
\begin{equation} \label{2e7}
G(x_1,x_2;y_1,y_2)_{\stackrel {{\displaystyle =}}{T\rightarrow \infty}}
\eta \sum_n \Phi_n(x_1,x_2)\overline \Phi_n(y_1,y_2)\ ,
\end{equation}
where $\eta =+1$ for bosonic fields and $\eta =-1$ for fermionic fields;
the sum over $n$ corresponds to a complete set of states; $\Phi_n$
is a generalized Bethe-Salpeter wave function with respect to the
intermediate state $|n>$ and becomes the usual Bethe-Salpeter wave
function when $|n>$ is a bound state \cite{sb,gml}. Although the function
$\Delta$ has a logarithmic increase at infinity, the charge neutrality
condition (\ref{2e5}) leads to a mutual cancellation of singular terms
and one obtains \cite{bcm}:
\begin{equation} \label{2e8}
\Phi_{\xi}(x_1,x_2)\ =\ e^{{\displaystyle ie^2\xi (\Delta(x_1-x_2) -
\Delta(0))}}\ \Phi(x_1,x_2)\ .
\end{equation}
\par
The infinite factor $\Delta(0)$ in Eqs. (\ref{2e2}), (\ref{2e4}) and
(\ref{2e8}) determines the transformation law of the wave function
renormalization constant $Z_2$ \cite{zjz}. {\it In the remaining part of
this work, we shall not consider radiative corrections and hence shall 
ignore gauge transformation effects coming from photons associated
entirely with one particle; only exchanged photon contributions will
be taken into account}. [More generally, one may consider different
gauges for exchanged photons and for photons entering in radiative
corrections \cite{by}.] Except in particular limiting procedures, such
as in Eq. (\ref{2e7}), there are no, in general, cancellations between
the two kinds of contributions, since they are concerned with different
variables.
\par
While the behavior under gauge transformations of Green's functions and 
wave functions is very simple (in $x$-space), the same is not true for
the inverses of Green's functions, the vertex functions and the scattering 
amplitudes. In this case, only infinitesimal gauge transformations have
relatively simple forms. They can be obtained either by inverting the
Green's functions or by using the Ward-Takahashi identities.
\par
The infinitesimal change of the inverse of the four-point Green's function
is:
\begin{equation} \label{2e9}
\delta_{\xi}G^{-1}\ =\ -G^{-1}(\delta_{\xi}G)G^{-1}\ .
\end{equation}
Equations (\ref{2e9}) and (\ref{2e4}) can then be used to show the
invariance, in a weak form, of the Bethe-Salpeter equation. The latter
can directly be written in terms of $G^{-1}$ as:
\begin{equation} \label{2e10}
G^{-1}\Phi\ =\ 0\ .
\end{equation}
Invariance of this equation under gauge transformations requires:
\begin{equation} \label{2e11}
(\delta_{\xi}G^{-1})\Phi + G^{-1}\delta_{\xi}\Phi\ =\ G^{-1}\left (
-(\delta_{\xi}G)G^{-1}\Phi + \delta_{\xi}\Phi\right )\ =\ 0\ .
\end{equation}
Replacing in this equation $\delta_{\xi}G$ and $\delta_{\xi}\Phi$ by
their expressions (\ref{2e4}) and (\ref{2e8}), respectively, one obtains:
\begin{eqnarray} \label{2e12}
& & \Delta(x_1-x_2) \int d^4z_1d^4z_2 G^{-1}(x_1,x_2;z_1,z_2) \Phi
(z_1,z_2)\nonumber \\
& & -\int d^4x_1'd^4x_2'd^4y_1d^4y_2d^4z_1d^4z_2 G^{-1}(x_1,x_2;
x_1',x_2')(\Delta(x_1'-y_2)+\Delta(x_2'-y_1))\nonumber \\
& & \times G(x_1',x_2';y_1,y_2)G^{-1}(y_1,y_2;z_1,z_2)\ \Phi(z_1,z_2)\ =
\ 0\ .
\end{eqnarray}
(The contribution of $\delta_{\xi}\Phi$ has been cancelled by one of the
terms of $\delta_{\xi}G$.)
The first term vanishes on account of Eq. (\ref{2e10}). In the second
and third terms, the gauge propagators $\Delta$ acting multiplicatively
on $G$, join one of the $x'$'s to the $y$ of the other line; in 
momentum space, they produce crossed diagrams with $G$; these do not 
have bound state poles at the positions of the direct diagram poles;
therefore, the two products $\Delta G$ cannot prevent the factor $G^{-1}
\Phi$ from vanishing, on account of Eq. (\ref{2e10}). One thus establishes
the weak gauge invariance of the Bethe-Salpeter equation.
\par
This conclusion is not changed when one includes radiative corrections
relative to fermion or boson lines and vertices. In particular, 
radiative corrections that join 
$x'$ to $y$ on the same line, imply in momentum space integrations on the
bound state pole position, which is then transformed into a cut.
\par
To display the gauge transformation property of the off-mass shell
scattering amplitude, we pass into momentum space. The infinitesimal
form of Eq. (\ref{2e4}) is:
\begin{eqnarray} \label{2e13}
& &\delta_{\xi}G(p_1,-p_2;p_1',-p_2')\ =\ ie^2\delta\xi \int \frac{d^4k}
{(2\pi)^4} \Delta(k)\big [G(p_1-k,-(p_2+k);p_1',-p_2') \nonumber \\
& &\ \ \ \ \ \ \ + G(p_1,-p_2,p_1'+k,-(p_2'-k))-G(p_1-k,-p_2;p_1',
-(p_2'-k)) \nonumber \\
& &\ \ \ \ \ \ \ - G(p_1,-(p_2+k);p_1'+k,-p_2')\big ]\ .
\end{eqnarray}
This equation is graphically represented in Fig. 1. We notice that if 
$G$ has a (simple) bound state pole, then in the right-hand side of Eq.
(\ref{2e13}) the first two terms have also the same pole. However, this
pole being simple, one concludes that the gauge transformation does not 
change the bound state energy \cite{by,ffh}. (Otherwise, the singularity
structure in the right-hand side of Eq. (\ref{2e13}) would be that of a
double pole.)
\par
The scattering amplitude $T$ is defined from the Green's function $G$ as:
\begin{equation} \label{2e14}
T\ =\ G_1^{-1}G_2^{-1}[G-G_0]G_1^{-1}G_2^{-1}\ ,
\end{equation}
where $G_1$ and $G_2$ are the external particle propagators and $G_0$
is their product:
\begin{equation} \label {2e15}
G_0\ =\ G_1 G_2\ .
\end{equation}
\par
The infinitesimal gauge transformation law of $T$ is then:
\begin{eqnarray} \label{2e16}
\delta_{\xi}T &=& ie^2\delta\xi \big[G_1^{-1}(p_1)-G_1^{-1}(p_1')\big]
\big[G_2^{-1}(-p_2)-G_2^{-1}(-p_2')\big]\Delta(p_1-p_1') \nonumber \\
& &\ +ie^2\delta\xi \int \frac{d^4k}{(2\pi)^4}\ \Delta(k)\ \big[\
G_1^{-1}(p_1)G_1(p_1-k) T(p_1-k,-(p_2+k);p_1',-p_2') \nonumber \\
& &\ \ \ \ \ \ \ \times G_2(-(p_2+k))G_2^{-1}(-p_2) \nonumber \\
& &\ \ \ +G_2^{-1}(-p_2')G_2(-(p_2'-k)) T(p_1,-p_2;p_1'+k,-(p_2'-k))
G_1(p_1'+k)G_1^{-1}(p_1') \nonumber \\
& &\ \ \ -G_1^{-1}(p_1)G_2^{-1}(-p_2')G_1(p_1-k)G_2(-(p_2'-k))
T(p_1-k,-p_2;p_1',-(p_2'-k)) \nonumber \\
& &\ \ \ -T(p_1,-(p_2+k);p_1'+k,-p_2') G_1(p_1'+k)G_2(-(p_2+k))
G_1^{-1}(p_1')G_2^{-1}(-p_2)\ \big]\ . \nonumber \\
& &
\end{eqnarray}
\par
Taking the external particles on their mass-shell, one immediately deduces
from this equation gauge invariance of the on-mass shell scattering 
amplitude.
\par
The unitarity of the gauge transformation can be checked with the 
invariance of the norm of the Bethe-Salpeter wave function. We first
notice that, according to Eq. (\ref{2e4}), the adjoint $\overline \Phi$
of the Bethe-Salpeter wave function [Eq. (\ref{2e7})] transforms in the
same way as $\Phi$ [Eq. (\ref{2e8})] (with the same sign in the 
exponential function); this is due to the fact that the gauge function 
$\Delta$ [Eq. (\ref{2e3})] is imaginary for euclidean variables in
$x$-space. The norm of the internal part of the Bethe-Salpeter wave
function is \cite{nnns}:
\begin{equation} \label{2e17}
(\phi,\phi)\ =\ \int d^4x\ \overline \phi\ \frac{\partial G^{-1}}
{\partial s}\ \phi\ =\ -i\eta\ ,
\end{equation}
where $s=P^2=(p_1+p_2)^2$ and $\eta$ was defined after Eq. (\ref{2e7}).
\par
Using infinitesimal gauge transformations together with Eqs. (\ref{2e8}),
(\ref{2e9}) and (\ref{2e13}) and restricting ourselves to hermitian
kernels and time-reversal invariant interactions, we find (in compact
notation):
\newpage
\begin{eqnarray} \label{2e18}
\delta_{\xi}(\phi,\phi) &=& ie^2\delta\xi \big \{ \int \overline \phi
\Delta\ \frac{\partial G^{-1}}{\partial s}\ \phi + \int \overline \phi
\ \frac{\partial G^{-1}}{\partial s}\ \Delta \phi \nonumber \\
& &\ -\int \overline \phi\ \frac{\partial G^{-1}}{\partial s}\ \big[
(\Delta G)+(G\Delta)-(\Delta G)_{cr}-(G\Delta)_{cr}\big]\ G^{-1}\phi
\nonumber \\
& &\ -\int \overline \phi G^{-1}\ \big[(\Delta G)+(G\Delta)-(\Delta
G)_{cr}-(G\Delta)_{cr}\big]\ \frac{\partial G^{-1}}{\partial s}\ \phi
\nonumber \\
& &\ -\int \overline \phi G^{-1}\ \frac{\partial}{\partial s}\big[
(\Delta G)+(G\Delta)-(\Delta G)_{cr}-(G\Delta)_{cr}\big]\ G^{-1}\phi
\big \}\ , 
\end{eqnarray}
where the subscript ``cr'' designates the terms corresponding to the
crossed diagrams [Eq. (\ref{2e13}) and Fig. 1]. In the above equation,
one is entitled to use the wave equation (\ref{2e10}), or its adjoint,
$\overline \phi G^{-1}=0$, as long as $G^{-1}$ is not multiplied by
terms having a pole at the bound state position in the $s$-channel.
Therefore, the crossed diagram terms disappear and after some algebraic
cancellations one remains with the following terms, which also
ultimately vanish:
\begin{eqnarray} \label{2e19}
\delta_{\xi}(\phi,\phi) &=& -ie^2\delta\xi\big\{\int \overline \phi\
\frac{\partial G^{-1}}{\partial s}\ (G\Delta)G^{-1}\phi + \int 
\overline \phi G^{-1}(\Delta G)\ \frac{\partial G^{-1}}{\partial s}\
\phi \nonumber \\
& &\ \ \ \ \ \ \ +\int \overline \phi G^{-1} \frac{\partial}{\partial s}
\big[\Delta G+G\Delta\big]G^{-1}\phi \big\} \nonumber \\
&=& -ie^2\delta\xi\big\{\int \overline \phi G^{-1}\frac{\partial}
{\partial s}\big(\Delta GG^{-1}\big)\phi + \int \overline \phi \frac
{\partial}{\partial s}\big(G^{-1}G\Delta\big)G^{-1}\phi\big\} \nonumber \\
&=& -ie^2\delta\xi\big\{\int \overline \phi \big(\frac{\partial \Delta}
{\partial s}\big)G^{-1}\phi + \int \overline \phi G^{-1}\big(\frac
{\partial \Delta}{\partial s}\big)\phi\big\} \nonumber \\
&=& 0\ .
\end{eqnarray}
\par
Let us finally remark that the gauge propagator $\Delta$, having a 
singular infra-red behavior, may lead, during the evaluation of
Feynman diagram integrals, to infra-red divergences. However, the
integrals involved in the variations $\delta_{\xi}G$ [Eq. (\ref{2e13})]
and $\delta_{\xi}T$ [Eq. (\ref{2e16})] are actually globally infra-red 
regular, although individual terms are divergent, as can be checked by
taking in the integrands the limit $k\rightarrow 0$. In later calculations,
some of the terms present in $\delta_{\xi}G$ or $\delta_{\xi}T$ will be
dropped, for they do not contribute to the bound state poles. It should
be understood, in such cases, that their infra-red divergent parts are kept,
in order to maintain the regularity of the remaining integral.
\par

\newpage

\section{Gauge transformations of constraint theory wave functions
and potentials}
\setcounter{equation}{0}

Constraint theory \cite{ct,cva1,s} allows the elimination of the 
relative energy variable of the two particles by means of a manifestly
covariant equation. The choice of the latter is not unique and
generally various choices are related one to the other by canonical
or wave function transformations. We choose the following constraint:
\begin{equation} \label{3e1}
[\ (p_1^2-p_2^2) - (m_1^2-m_2^2)\ ]\ \widetilde \Psi (x_1,x_2)\ =\ 0\ ,
\end{equation}
where $p_1$ and $p_2$ are the momentum operators of particles 1 and 2,
$m_1$ and $m_2$ their respective masses and $\widetilde \Psi$ is the
constraint theory wave function.
We use standard definitions for the total and relative variables:
$P= p_1 + p_2$, $p = (p_1-p_2)/2$, $x=x_1-x_2$, $M=m_1 + m_2$. 
For states that are eigenstates of the total momentum $P$, we define
transverse and longitudinal components of four-vectors with respect
to $P$ and denote them with indices $T$ and $L$, respectively; thus:
$x_{\mu}^T=x_{\mu}-x.\hat P \hat P_{\mu}$, $x_L=x.\hat P$,
$\hat P_{\mu} = P_{\mu}/\sqrt {P^2}$, $P_L = \sqrt {P^2}$,
$r=\sqrt {-x^{T2}}$.
\par
With these definitions, constraint (\ref{3e1}) can be written in the
following form:
\begin{equation} \label{3e4}
C(p)\ \equiv \ 2P_Lp_L - (m_1^2-m_2^2)\ \approx \ 0\ .
\end{equation}
Constraint $C$ allows the elimination of the relative longitudinal
momentum $p_L$ and fixes the evolution law with respect to its conjugate
variable $x_L$. The internal dynamics of the system becomes 
three-dimensional, apart from the spin degrees of freedom, expressed
in terms of the transverse vector $x^T$.
\par
In the presence of constraint $C$, the individual Klein-Gordon operators
become equal:
\begin{equation} \label{3e5}
H_0\ \equiv \ (p_1^2-m_1^2)\big \vert _C \ =\ (p_2^2-m_2^2)\big
\vert _C \ =\ \frac {P^2}{4} - \frac {1}{2} (m_1^2+m_2^2) +
\frac {(m_1^2-m_2^2)^2}{4P^2} + p^{T2}\ .
\end{equation}
\par
The wave equations for two spin-0 particle systems are:
\subequations
\begin{eqnarray}
\label{3e6a}
(p_1^2 - m_1^2 - \widetilde V)\widetilde \Psi (x_1,x_2) &=& 0\ ,\\
\label{3e6b}
(p_2^2 - m_2^2 - \widetilde V)\widetilde \Psi (x_1,x_2) &=& 0\ ,  
\end{eqnarray}
\endsubequations
while for spin-$\frac{1}{2}$ fermion-antifermion systems they are
\cite{s,ms}:
\subequations
\begin{eqnarray}
\label{3e7a}
(\gamma _1.p_1-m_1)\widetilde \Psi &=& (-\gamma _2.p_2+m_2)
\widetilde V \widetilde \Psi\ ,\\
\label{3e7b}
(-\gamma _2.p_2-m_2)\widetilde \Psi &=& (\gamma _1.p_1+m_1)
\widetilde V \widetilde \Psi\ .
\end{eqnarray}
\endsubequations              
[The antifermion Dirac matrices $\gamma_2$ act on $\widetilde \Psi$,
represented as a $4\times 4$ matrix, from the right, in the reverse
order of their appearance.] Potentials $\widetilde V$ are Poincar\'e
invariant operators and depend on $x$ only through the transverse
vector $x^T$.
\par
Equations (\ref{3e6a})-(\ref{3e6b}) and (\ref{3e7a})-(\ref{3e7b})
can also be written in a unified form. Introducing the individual
particle propagators $G_1$ and $G_2$ (with $i$-factors in the 
numerators in momentum space) and their product $G_0$ [Eq. (\ref{2e15})]
and defining
\begin{equation} \label{3e8}
\widetilde g_0\ =\ H_0G_0\big\vert_{C(p)}\ ,
\end{equation}
Eqs. (\ref{3e6a})-(\ref{3e6b}) and (\ref{3e7a})-(\ref{3e7b}) take the
form:
\begin{equation} \label{3e9}
(\widetilde g_0^{-1}+\widetilde V)\ \widetilde \Psi\ =\ 0\ .
\end{equation}
\par
In order to establish the connection of constraint theory wave equations
and potentials with the Bethe-Salpeter equation and related quantities,
one projects, with an appropriate weight factor, which is chosen here
to be $H_0$ [Eq. (\ref{3e5})], Green's functions, scattering amplitudes
and wave functions on the constraint hypersurface (\ref{3e4}). Thus,
defining the left-projected Green's function $\widetilde G$,
\begin{equation} \label{3e16}
\widetilde G(P,p,p')\ =\ -2i\pi \delta(C(p)) H_0 G(P,p,p')\ ,
\end{equation}
one can iterate, in the right-hand side, $G$ around $\widetilde G$,
repeatedly using its integral equation. One ends up with an integral
equation satisfied by $\widetilde G$, the kernel of which is related
to the Bethe-Salpeter kernel $K$ \cite{js}. Defining the constraint 
theory wave function $\Psi_C$ from the residue of $\widetilde G$ at
a bound state pole,
\begin{equation} \label{3e18}
\Psi_C\ \equiv\ 2\pi 2P_L \delta(C)\widetilde \Psi\ 
=\ -2i\pi \delta(C) H_0 \Phi\ ,
\end{equation}
one finds that the wave function $\widetilde  \Psi$ 
satisfies Eq. (\ref{3e9}) with $\widetilde V$ related to the scattering
amplitude by means of a Lippmann-Schwinger-quasipotential type
equation \cite{qp,lcl,js}:
\begin{eqnarray} 
\label{3e19}
& &\widetilde V\ =\ \widetilde T (1-\widetilde g_0\widetilde T)^{-1}\ ,\\
\label{3e20}
& &\widetilde T(p,p')\ =\ \frac {i}{2P_L} T(p,p')\big \vert _{C(p),C(p')}\ .
\end{eqnarray}  
[In $T$, the total four-momentum conservation factor $(2\pi)^4 \delta^4
(P-P')$ has been amputated.]
\par
Conversely, the Bethe-Salpeter wave function $\Phi$ can be reconstructed
from $\Psi_C$ with the equation
\begin{equation} \label{3e21}
\Phi\ =\ G_0T(1-\widetilde gT)^{-1} \Psi_C\ ,\ \ \ \ \ \ \ 
\widetilde g\ =\ 2i\pi \delta(C)\widetilde g_0\ ,
\end{equation}
$G_0$ and $\widetilde g_0$ being defined in Eqs. (\ref{2e15}) and
(\ref{3e8}).
\par
Equation (\ref{3e19}) is the basis for the calculation of the potential
of constraint theory from the scattering amplitude. Iterating Eq. 
(\ref{3e19}) with respect to $\widetilde T$, one finds that $\widetilde V$
receives contributions, in addition to those of the ordinary Feynman 
diagrams, from new diagrams having at least one three-dimensional box
sub-diagram, corresponding to the presence of the constraint factor
$\widetilde g_0$. These diagrams, which we call ``constraint diagrams'',
play a crucial role in the cancellation mechanism of spurious infra-red
singularities \cite{rta,js} and in the reorganiszation of the 
perturbation series.
\par
The infinitesimal gauge transformation law of the constraint theory
wave function is obtained by starting from Eqs. (\ref{3e18}) and using
Eqs. (\ref{3e21}) and (\ref{2e8}). One has:
\begin{eqnarray} \label{4e6}
\delta_{\xi}\widetilde \Psi &=& -\frac{i}{2P_L} H_0
\delta_{\xi}\Phi\big\vert_C \nonumber \\
&=& -\frac{i}{2P_L}ie^2\delta\xi H_0 (\Delta \Phi)\big\vert_C \nonumber \\
&=& -\frac{i}{2P_L}ie^2\delta\xi H_0\ (\Delta G_0T)\ (1-\widetilde g_0
\widetilde T)^{-1}\widetilde \Psi\ ,
\end{eqnarray}
where the integration inside the term $(\Delta G_0T)$ is
four-dimensional.
\par
The transformation law of the potential is obtained by starting from the
relationship between the wave equation operator $(\widetilde g_0^{-1}
+\widetilde V)$ [Eq. (\ref{3e9})] and the scattering amplitude
$\widetilde T$ [Eqs. (\ref{3e19})-(\ref{3e20})]:
\begin{equation} \label{4e7}
(\widetilde g_0^{-1}+\widetilde V)\ =\ \widetilde g_0^{-1} (1-
\widetilde g_0\widetilde T)^{-1}\ =\ (1-\widetilde T\widetilde g_0)
^{-1}\widetilde g_0^{-1}\ .
\end{equation}
One finds the relation
\begin{equation} \label{4e7p}
\delta_{\xi}(\widetilde g_0^{-1}+\widetilde V)\ =\ (\widetilde g_0^{-1}
+\widetilde V)\widetilde g_0(\delta_{\xi} \widetilde T)\widetilde g_0
(\widetilde g_0^{-1}+\widetilde V)\ ,
\end{equation}
which, according to the transformation law (\ref{2e16}), becomes:
\begin{eqnarray} \label{4e8}
\delta_{\xi}(\widetilde g_0^{-1}+\widetilde V) &=& \frac{i}{2P_L}
ie^2\delta\xi(\widetilde g_0^{-1}+\widetilde V)\widetilde g_0\big\{
G_0^{-1}(\Delta + \Delta G_0T) \nonumber \\
& &\ \ \ \ +(\Delta + TG_0\Delta)G_0^{-1} - ({\rm crossed})\big\}
\widetilde g_0(\widetilde g_0^{-1}+\widetilde V)\ .
\end{eqnarray}
\par
When this equation is applied on the wave function $\widetilde \Psi$,
the operator $(\widetilde g_0^{-1}+\widetilde V)$ of the utmost right
gives zero [Eq. (\ref{3e9})] provided it is not multiplied by terms
having the bound state pole at the same position. Therfore, the crossed
terms, as well as the single $\Delta$ terms, disappear from the equation.
Furthermore, in the second term of the right-hand side of the equation 
the product $(i/(2P_L))(\widetilde g_0^{-1}+\widetilde V)\widetilde g_0
T$ is , according to Eqs. (\ref{4e7}), the four-dimensional continuation 
(in relative longitudinal momentum) of $(1-\widetilde T\widetilde g_0)
^{-1}\widetilde T=\widetilde V$, which does not have a pole; hence, it does 
not contribute when Eq. (\ref{4e8}) is applied on $\widetilde \Psi$.
One thus obtains the equation:
\begin{equation} \label{4e9}
[\delta_{\xi}(\widetilde g_0^{-1}+\widetilde V)]\ \widetilde \Psi\ =\
\frac{i}{2P_L} ie^2\delta\xi(\widetilde g_0^{-1}+\widetilde V) H_0
(\Delta G_0T)(1-\widetilde g_0\widetilde T)^{-1}\ \widetilde \Psi\ ,
\end{equation}
where Eq. (\ref{3e8}) was used.
\par
The combination of the two transformation laws (\ref{4e6}) and (\ref{4e9})
leads to the weak invariance of the wave equation (\ref{3e9}):
\begin{equation} \label{4e10}
\delta_{\xi}[\ (\widetilde g_0^{-1}+\widetilde V) \widetilde \Psi\ ]\
\approx\ 0\ .
\end{equation}
\par
The invariance of the norm of $\widetilde \Psi$ can be shown in a similar
way as for the Bethe-Salpeter wave function [Eqs. 
(\ref{2e17})-(\ref{2e19})]. The norm of the internal part of $\widetilde 
\Psi$ is \cite{lcl,s,js}:
\begin{equation} \label{4e11}
(\widetilde \psi,\widetilde \psi)\ =\ -\eta \int d^3x^T 4P^2\ \overline
{\widetilde \Psi}\ \frac{\partial}{\partial s}[\widetilde g_0^{-1}+
\widetilde V]\ \widetilde \psi\ =\ 2P_L\ ,
\end{equation}
where $\overline {\widetilde \psi}$ is the adjoint of $\widetilde \psi$,
equal to $\widetilde \psi^*$ in the bosonic case and to $[\gamma_{1L}
\gamma_{2L}\widetilde \psi]^{\dagger}$ in the fermionic case; $s=P^2$
and $\eta$ was defined after Eq. (\ref{2e7}). Using the transformation
laws (\ref{4e6}) and (\ref{4e8}) and arguments similar to those used in
Eqs. (\ref{2e17})-(\ref{2e19}) one shows the (weak) invariance of the 
norm (\ref{4e11}):
\begin{equation} \label{4e12}
\delta_{\xi}(\widetilde \psi,\widetilde \psi)\ =\ 0\ .
\end{equation}
\par
Contrary to the four-dimensional case, the gauge transformation laws 
(\ref{4e6}) and (\ref{4e9}) of the three-dimensional theory explicitly 
depend on the interaction, a feature that renders their evaluations
rather tricky. Furthermore, the amplitude T that appears in the 
transformation is not submitted on its left to the constraint (\ref{3e4}).
Therefore, the quantity $(i/(2P_L))T(1-\widetilde g_0\widetilde T)^{-1}$
represents a four-dimensional continuation of potential $\widetilde V$
[Eq. (\ref{3e19})]. Its evaluation, in the general case, cannot be done in
compact form. However, when the local approximation is used for the
potentials, simplifications occur. This case is considered in Sec. 4.
\par

\newpage

\section{Gauge transformations in the local \protect \\ approximation 
of potentials}
\setcounter{equation}{0}

In lowest order of perturbation theory (one photon exchange),
relationship (\ref{3e19}) provides a local expression for $\widetilde V$
in three-dimensional $x$-space (with respect to $x^T$). It turns out
that this property can also be maintained, in a certain 
approximation, in higher orders. The perturbation theory calculations
effected in Ref. \cite{js} have shown that, in the Feynman gauge, the
infra-red leading part of the contribution of the $n$-photon exchange 
diagrams can be represented in (three-dimensional) $x$-space by a local
function of $r$ ($=\sqrt{-x^{T2}}$), of the type $(\alpha/r)^n$, $\alpha$
being the fine structure constant. The sum of these leading terms also
yields a local function for $\widetilde V$. Therefore, the local 
approximation of $\widetilde V$ can be considered as a sensible one: it
includes not only lowest order effects, but also leading effects of 
multiphoton exchange diagrams. The use of this approximation
considerably simplifies the resolution of wave equations, where now 
standard methods of quantum mechanics can be applied. In the rest of this
article we shall limit ourselves to this approximation and shall 
correspondingly consider the transformation laws obtained in Sec. 3. 
\par
In the fermionic case, to ensure positivity of the norm, potential 
$\widetilde V$ must satisfy an inequality \cite{ms}; for local potentials
that commute with $\gamma_{1L}\gamma_{2L}$, the parametrization \cite{cva2}
\begin{equation} \label{3e10}
\widetilde V\ =\ \tanh V\ ,
\end{equation}
satisfies this condition. Introducing the wave function transformation
\begin{equation} \label{3e11}
\widetilde \Psi\ =\ (\cosh V)\Psi\ ,
\end{equation}
the norm of the internal part of the new wave function $\Psi$ becomes
(in the c.m. frame):
\begin{equation} \label{3e12}
\int d^3{\bf x}\ Tr \bigg \{ \psi ^{\dagger}\big [ 1+4\gamma _{10}
\gamma _{20} P_0^2 \frac {\partial V}{\partial P^2}\big ]\psi
\bigg \}\ =\ 2P_0\ ,
\end{equation}
which, for c.m. energy independent potentials, reduces to the 
conventional free norm of states.
\par
Considering local potentials $V$ (in $x^T$) composed of combinations
of scalar, pseudoscalar and vector potentials,
\begin{equation} \label{3e13}
V\ =\ V_1 + \gamma _{15}\gamma _{25}V_3 + \gamma _1^{\mu}\gamma _2^
{\nu} (g_{\mu\nu}^{LL}V_2 + g_{\mu\nu}^{TT}U_4 + \frac {x_{\mu}^T
x_{\nu}^T}{x^{T2}}T_4)\ ,
\end{equation}
and using transformations (\ref{3e10})-(\ref{3e11}), Eqs. 
(\ref{3e7a})-(\ref{3e7b}) can be brought into forms analogous to the
Dirac equation, where each particle appears as placed in the external
potential created by the other particle. The wave equation satisfied
by particle 1 becomes \cite{ms}:
\begin{eqnarray} \label{3e14}
& &\bigg \{ \big [\frac {P_L}{2} e^{\displaystyle 2V_2} + 
\frac {(m_1^2 - m_2^2)}{2P_L}
e^{\displaystyle -2V_2}\ \big ]\ \gamma _{1L}  - 
\frac {M}{2} e^{\displaystyle 2V_1}
- \frac {(m_1^2 - m_2^2)}{2M} e^{\displaystyle -2V_1} \nonumber \\
& &\ +e^{\displaystyle -2U_4}\bigg[\gamma _1^T.p^T + \frac {i\hbar}{2x^{T2}}
(e^{\displaystyle -2T_4} - 1)\ (2\gamma _1^T.x^T + 
i\gamma _1^{T\alpha } \sigma _{2\alpha \beta } ^{TT} x^{T\beta }) 
+ (e^{\displaystyle -2T_4} - 1) \frac {\gamma _1^T.x^T}{x^{T2}} 
x^T.p^T\nonumber \\
& &\ - 2i\hbar e^{\displaystyle -2T_4} \gamma _2^T.x^T
\big (\dot V_1 + \gamma _{1L}
\gamma _{2L} \dot V_2 + \gamma _{15} \gamma _{25} \dot V_3 + \gamma _1^T.
\gamma _2^T \dot U_4+\frac{\gamma _1^T.x^T \gamma _2^T.x^T}{x^{T2}}\dot T_4
\big ) \bigg ] \bigg \} \Psi \ =\ 0\ ,\nonumber \\
& & 
\end{eqnarray}
where we have defined:
\begin{equation} \label{3e15}
\dot F \ \equiv \ \frac {\partial F}{\partial x^{T2}}\ ,\ \ \ \ \ \
\sigma _{a\mu\nu} = \frac {1}{2i} \big [ \gamma _{a\mu},
\gamma _{a\nu} \big ]\ \ \ (a=1,2)\ .
\end{equation}
The wave equation satisfied by particle 2 can be obtained from Eq. 
(\ref{3e14}) by the
replacements : $p_1 \leftrightarrow -p_2$, $x \rightarrow x$, $m_1
\leftrightarrow m_2$, $\gamma _1 \leftrightarrow \gamma _2$ . 
We recognize that the scalar potential $V_1$ acts as a 
modification of the total mass $M$ of the fermions through the change 
$M\rightarrow Me^{{\displaystyle 2V_1}}$ while $(m_1^2-m_2^2)$ is kept 
fixed. The timelike vector potential $V_2$ acts as a modification of the 
c.m. total energy $P_L$ through the change $P_L\rightarrow P_L
e^{{\displaystyle 2V_2}}$, while $(p_{1L}^2-p_{2L}^2)=(m_1^2-m_2^2)$ is 
kept fixed. The spacelike potential $U_4$ changes the orbital 
angular momentum operator from ${\bf L}$ to 
${\bf L}e^{{\displaystyle -2U_4}}$ (in the c.m. frame) and the combination 
$U_4+T_4$ of the
spacelike potentials changes the radial momentum operator from $p_r$ to 
$p_r e^{{\displaystyle -2(U_4+T_4)}}$ (in the classical limit). The 
pseudoscalar potential appears only in spin- and $\hbar$-dependent terms.
\par
In QED, the summation of leading infra-red effects of multiphoton
exchange diagrams, as described above \cite{js}, leads to the following
expressions for the timelike ($V_2$) and spacelike ($U_4$ and $T_4$)
parts of the electromagnetic potential in the Feynman gauge:
\begin{eqnarray}
\label{3e23}
V_2 &=& \frac{1}{4} \ln \left (1+\frac{2\alpha}{P_L r}\right )\ , 
\\
\label{3e24}
U_4 &=& V_2\ ,\ \ \ \ \ T_4\ =\ 0\ .
\end{eqnarray}
\par
Potentials (\ref{3e23})-(\ref{3e24}) are compatible with the minimal
substitution rules proposed long ago by Todorov for spin-0 particles,
on the basis of an identification of the two-particle motion in the c.m.
frame to that of a fictitious particle with appropriately defined reduced
mass and energy \cite{t2}.These rules were extended to the fermionic
case by Crater and Van Alstine \cite{cva1}. The above potentials were
shown to reproduce the correct $O(\alpha^4)$ effects in muonium and
positronium spectra \cite{cbwva,ms}.
\par
Similar results as above can also be derived with scalar photons contributing
to the scalar potential $V_1$ [Eq. (\ref{3e13})] \cite{cva1,js}.
\par
The application of the previous summation method of the Feynman diagrams 
to the intercations of bosons shows, as one naturally expects, that the
classical parts of the potentials in the fermionic case (written in the
Pauli-Schr\"odinger form \cite{ms}) and in the bosonic case coincide
\cite{js}. Therefore, one can use unified potentials for both cases.
\par
It is not straightforward to generalize the above evaluation and
summation techniques to other covariant gauges than the Feynman gauge.
The presence of the additional gauge piece in the photon propagator
breaks the permutational symmetry used in the previous calculations
and renders their evaluation rather complicated. This is why these
potentials will be evaluated from the Feynman gauge, using the infinitesimal
transformation laws obtained in Sec. 3.
\par
To have a rough idea of the expected results, we first consider in some 
detail the one-photon exchange approximation . The corresponding 
potential is then local, without further approximation. Indeed, Eq.
(\ref{3e19}), specialized to the one-photon exchange diagram, projects,
with the constraint condition (\ref{3e4}), the photon propagator in 
momentum space on the surface $k_L=0$; the potential in $x$-space is
obtained with the three-dimensional Fourier transformation with respect to
$k^T$. One finds for the photon propagator in three-dimensional $x$-space,
in the gauge $\xi$, the expression:
\begin{equation} \label{4e1}
\widetilde D_{\mu\nu}(x^T)\ =\ \frac{i}{4\pi}\left ( g_{\mu\nu}^{LL}
+ g_{\mu\nu}^{TT}(1-\frac{\xi}{2}) + \frac{x_{\mu}^Tx_{\nu}^T}{x^{T2}}
\frac{\xi}{2}\right ) \frac{1}{r}\ ;
\end{equation}
it yields the following expressions for the potentials in the gauge $\xi$:
\begin{equation} \label{4e2}
V_{2\xi}\ =\ \frac{\alpha}{2P_L r}\ ,\ \ \ \ \ \ U_{4\xi}\ =\ 
V_{2\xi}+U_{g\xi}\ ,\ \ \ \ \ \ T_{4\xi}\ =\ 2x^{T2}\dot U_{g\xi}\ ,
\ \ \ \ \ \ U_{g\xi}\ =\ -\frac{\xi}{2}\ \frac{\alpha}{2P_L r}\ .
\end{equation}
[The dot operation is defined in Eq. (\ref{3e15}).] To this order,
$V_{2\xi}$ is independent of $\xi$.
\par
After replacing these potentials in the wave equation (\ref{3e14}) 
(and in the equivalent one of particle 2) and designating by $\Psi_{\xi}$
the corresponding wave function, it can be seen that the wave function
transformation 
\begin{eqnarray} \label{4e3}
\Psi_{\xi} &=& U(\xi) \Psi\ , \ \ \ \ \ \ 
U(\xi)\ \simeq\ (1+ie^2\xi S_0)\ ,\nonumber \\
S_0 &=& \frac{1}{2} (Fx^T.p^T+p^T.x^T F)\ ,\ \ \ \ \ F=\frac{1}{4\pi}
\ \frac{1}{2P_L r}\ ,
\end{eqnarray}
removes, to first order in $\alpha$, all the $\xi$-dependent terms from
the wave equation and gives back the wave equation in the Feynman gauge.
The operator $S_0$ can also be written in the following form:
\begin{equation} \label{4e4}
S_0\ =\ -\frac{i}{4}\ \left [ H_0, \int^{x^{T2}} dx^{T2} F \right ]
\ =\ -\frac{i}{2P_L} \left [ H_0, \widetilde \Delta (x^T) \right ]\ ,
\end{equation}
where $\widetilde \Delta (x^T)$ is the three-dimensionally reduced 
expression of the gauge propagator $\Delta(x)$ [Eq. (\ref{2e3})]
(calculated by the Schwinger parametrization and dimensional
regularization):
\begin{equation} \label{4e5}
\widetilde \Delta (x^T)\ =\ \int \frac{d^3k^T}{(2\pi)^3}\ \frac
{e^{{\displaystyle ik^T.x^T}}}{(k^{T2}+i\epsilon)^2}\ =\ -\frac{1}{8\pi}
r\ .
\end{equation}
\par
The above study can also be repeated for the bosonic case, the same
results as in Eqs. (\ref{4e3})-(\ref{4e4}) being found. [The wave
equations for vector interactions with bosons can be found in Refs.
\cite{cva1,s}. In the first paper of Ref. \cite{s}, the eigenvalue
equation in the Feynman gauge is given by Eq. (5.12) with the 
identifications $(1-B)=(1-A)^{-1}=e^{{\displaystyle 2V_2}}=
e^{{\displaystyle 2U_4}}$.]
\par
To investigate the transformation laws for the higher order diagrams,
we go back to the general case of Eqs. (\ref{4e6}) and (\ref{4e9}).
We use the wave equation (\ref{3e9}), together with relation (\ref{3e19}),
in its integral form, valid for a bound state,
\begin{equation} \label{5e1}
\widetilde \Psi + \widetilde g_0\widetilde T (1-\widetilde g_0\widetilde
T)^{-1} \widetilde \Psi\ =\ 0\ ,
\end{equation}
and add it, multiplied with an appropriate (nonsingular) factor, to Eq.
(\ref{4e6}). We obtain:
\begin{equation} \label{5e2}
\delta_{\xi}\widetilde \Psi\ =\ -\frac{i}{2P_L} ie^2\delta\xi H_0
\left \{ (\Delta G_0T)+(\widetilde \Delta \widetilde g_0\widetilde T)
\right \} (1-\widetilde g_0\widetilde T)^{-1} \widetilde \Psi 
-\frac{i}{2P_L} ie^2\delta\xi H_0\widetilde \Delta \widetilde \Psi\ .
\end{equation}
The quantity $(\widetilde \Delta \widetilde g_0\widetilde T)$ is the 
constraint diagram counterpart of the amplitude $(\Delta G_0T)$ \cite{js} 
and the integration inside it is three-dimensional, after constraint 
(\ref{3e4}) is used.
\par
We can still add to Eq. (\ref{5e2}) the contribution of the crossed
diagram counterpart of $(\Delta G_0T)$, denoted by $(\Delta G_0T)_{cr}$,
in which the gauge propagator $\Delta$ crosses the scattering amplitude
$T$ (see Fig. 2). This is possible since $(\Delta G_0T)_{cr}$ does not
have a pole at the bound state position in the $s$-channel and hence one 
can apply the operator $(1-\widetilde g_0\widetilde T)^{-1}$ on 
$\widetilde \Psi$ and obtain zero. Thus, Eq. (\ref{5e2}) becomes:
\begin{eqnarray} \label{5e3}
\delta_{\xi}\widetilde \Psi &=& -\frac{i}{2P_L} ie^2\delta\xi H_0
\left \{ (\Delta G_0T)+(\Delta G_0T)_{cr}+(\widetilde \Delta \widetilde 
g_0\widetilde T) \right \} (1-\widetilde g_0\widetilde T)^{-1}\widetilde 
\Psi \nonumber \\
& & \ \ \ \ 
-\frac{i}{2P_L} ie^2\delta \xi H_0\widetilde \Delta \widetilde \Psi\ .
\end{eqnarray}
\par
The sum of the amplitudes $(\Delta G_0T)$, $(\Delta G_0T)_{cr}$ and
$(\widetilde \Delta \widetilde g_0\widetilde T)$ can be evaluated at
leading order of the infra-red counting rules of QED with the eikonal
approximation \cite{cw,lesu,aibizj} adapted to the bound state problem.
This approximation was verified to yield the correct results for the
leading terms of the two-photon exchange diagrams  and then was
generalized to higher-order diagrams \cite{js}. It consists of making
the following approximations in the fermion propagators:
\begin{eqnarray} \label{5e4}
G_1(p_1-k_1) &\simeq& \frac{i}{-2p_1.k_1+i\epsilon}\big [(\gamma_{1L}
p_{1L}+m_1)-\gamma_{1L}k_{1L}\big ]\ ,\nonumber \\
G_2(-(p_2+k_2)) &\simeq& \frac{i}{2p_2.k_2+i\epsilon} \big [(-\gamma_{2L}
p_{2L}+m_2)-\gamma_{2L}k_{2L}\big ]\ ,
\end{eqnarray}
and of neglecting, at intermediate stages of the calculation,
momentum transfers relative to subgraphs of a given graph. 
Neglecting thus in the amplitude $T$, in Eq. (\ref{5e3}), the momentum
transfer, the calculation becomes similar to that of two-photon exchange
diagrams. (For positivity reasons, we also retain the quadratic term
$k_L^2$ in the product $G_1G_2$.) One finds: 
\begin{equation} \label{5e5}
(\Delta G_0T)+(\Delta G_0T)_{cr}+(\widetilde \Delta \widetilde g_0
\widetilde T)\ \simeq\ \widetilde \Delta (2+\frac{H_0}{4p_{1L}p_{2L}})
\widetilde T\ .
\end{equation}
At leading order, the term $H_0/(4p_{1L}p_{2L})$ is equivalent to 
$-\widetilde g_0^{-1}$; neglecting quantum effects, the latter can be
brought to the utmost right and replaced there by $\widetilde V$. One
finally obtains:
\begin{equation} \label{5e6}
\delta_{\xi}\widetilde \Psi\ =\ -\frac{i}{2P_L} ie^2\delta \xi H_0
\widetilde \Delta (1+\widetilde V_{lead,\xi})^2 \widetilde \Psi\ ,
\end{equation}
where $\widetilde V_{lead,\xi}$ is the infra-red leading part of 
$\widetilde V$ (in the gauge $\xi$), 
i.e., the timelike component of the vector potential and where 
$\gamma_{1L}\gamma_{2L}$ is replaced by $-1$:
\begin{equation} \label{5e7}
\widetilde V_{lead,\xi}\ =\ -\tanh V_{2\xi}\ .
\end{equation}
(In the Feynman gauge $V_2$ is given by Eq. (\ref{3e23}).) 
\par
Similarly, Eq. (\ref{4e9}) yields:
\begin{equation} \label{5e8}
[\delta_{\xi}(\widetilde g_0^{-1}+\widetilde V)]\widetilde \Psi\ =\
\frac{i}{2P_L}ie^2\delta\xi (\widetilde g_0^{-1}+\widetilde V) H_0
\widetilde \Delta (1+\widetilde V_{lead,\xi})^2 \widetilde \Psi\ .
\end{equation}
\par
To integrate Eq. (\ref{5e6}) up to finite $\xi$'s, we bring the
operator $H_0$ to the utmost right and use the equation of motion (with
the approximation $H_0\simeq -4p_{1L}p_{2L}\widetilde g_0^{-1}$):
\begin{equation} \label{5e9}
\delta_{\xi}\widetilde \Psi\ =\ -\frac{i}{2P_L}ie^2\delta \xi \big \{
\big[H_0,\widetilde \Delta (1+\widetilde V_{lead,\xi})^2\big] + 4p_{1L}
p_{2L}\widetilde \Delta (1+\widetilde V_{lead,\xi})^2\widetilde V_{lead,\xi}
\big \}\ \widetilde \Psi\ .
\end{equation}
The solution of this equation is:
\begin{eqnarray} \label{5e10}
& & \widetilde \Psi_{\xi_2}\ =\ \overline U(\xi_2,\xi_1)\widetilde 
\Psi_{\xi_1}\ , \nonumber \\
& & \overline U(\xi_2,\xi_1)\ =\ {\cal P}\left(\exp\big\{ie^2
\int_{\xi_1}^{\xi_2} d\xi W(\xi)\big\}\right)\ , \nonumber \\
& & W(\xi)\ =\ -\frac{i}{2P_L}\big\{\big[H_0,\widetilde \Delta (1+
\widetilde V_{lead,\xi})^2\big] + 4p_{1L}p_{2L}\widetilde \Delta (1+
\widetilde V_{lead,\xi})^2\widetilde V_{lead,\xi}\big\}\ ,
\end{eqnarray}
where ${\cal P}$ is the path ordering operator.
\par
In the following, we shall study an approximate form of this 
transformation law. To this end, we adopt two simplifications. First, 
we assume that in the commutator, in $W(\xi)$, the potential
$\widetilde V_{lead,\xi}$ can be approximated by its expression of the
Feynman gauge:
\begin{equation} \label{5e10p}
\widetilde V_{lead,\xi}\ \simeq\ \widetilde V_{lead,F}\ =\ 
\left(\frac{1-\sqrt{1+\frac{2\alpha}{P_L r}}}{1+\sqrt{1+\frac{2\alpha}
{P_L r}}}\right)\ .
\end{equation}
Second, we assume that the two operators in $W(\xi)$ are commuting
objects. With these approximations the gauge transformation operator
$\overline U(\xi_2,\xi_1)$ takes the form:
\begin{eqnarray} \label{5e11}
& & \overline U(\xi_2,\xi_1)\ \simeq\ T(\xi_2,\xi_1) U(\xi_2-\xi_1)\ ,
\nonumber \\
& & U(\xi)\ =\ e^{{\displaystyle ie^2\xi S}}\ ,\ \ \ \ \ \ \ \ 
S\ =\ -\frac{i}{2P_L}\big[H_0,\widetilde \Delta(1+\widetilde V_{lead,F})
^2\big]\ , \nonumber \\
& & T(\xi_2,\xi_1)\ =\ \exp\big\{\frac{e^2}{2P_L}\int_{\xi_1}^{\xi_2} 
d\xi\ 4p_{1L}p_{2L}\widetilde \Delta (1+\widetilde V_{lead,\xi})^2
\widetilde V_{lead,\xi}\big\}\ .
\end{eqnarray}
($T(\xi_2,\xi_1)$ and $U(\xi)$ are supposed to be commuting.)
As we shall see in Sec. 5, the above approximate forms provide the main
qualitative properties of the gauge transformations of wave functions
and potentials.
\par
The wave equation operator $(\widetilde g_0^{-1}+\widetilde V)$ 
transforms as:
\begin{equation} \label{5e12}
[\widetilde g_0^{-1}+\widetilde V]_{\xi}\ =\ T^{-1}(\xi_2,\xi_1)
U(\xi_2-\xi_1)(\widetilde g_0^{-1}+\widetilde V)U^{\dagger}(\xi_2-\xi_1)
T^{-1}(\xi_2,\xi_1)\ .
\end{equation}
Transformations (\ref{5e11})-(\ref{5e12}) ensure the (weak) invariance
of the norm (\ref{4e11}).
\par
Among the two transformation operators $U$ and $T$, it is the former 
which is the nontrivial one, generating local transformations in 
$x$-space, while the latter acts as a multiplicative factor. In the
following we shall focus our attention on the properties of the 
operator $U$.
\par

\newpage

\section{Properties of gauge transformations in the local approximation
of potentials}
\setcounter{equation}{0}

This section is devoted to the study of the properties of gauge
transformations in the local approximation of potentials, implemented
by the operator $U$ [Eq. (\ref{5e11})]. There is a complete similarity 
between the cases of bosons and fermions (the operator $U$ is spin
independent) and for this reason we shall concentrate on
the case of fermions only. The generator $S$ of the transformations will 
be written in a form similar to that of Eq. (\ref{4e3}), which is more
tractable for practical calculations. Thus, the gauge transformation
operator $U(\xi)$ is defined as:
\begin{eqnarray} \label{6e1}
& & \widetilde \Psi_{\xi}\ =\ U(\xi) \widetilde \Psi\ ,\ \ \ \ \ \ \
U(\xi)\ =\ e^{{\displaystyle i\xi(fx^T.p^T+p^T.x^Tf)/(2\hbar)}}\ ,
\nonumber \\
& & f\ =\ \frac{\alpha}{2P_L r}\ \frac{\partial}{\partial r} [r(1+
\widetilde V_{lead,F})^2]\ ,\ \ \ \ \ \widetilde V_{lead,F}\ =\ \left (
\frac{1-\sqrt{1+\frac{2\alpha}{P_L r}}}{1+\sqrt{1+\frac{2\alpha}
{P_L r}}}\right )\ .
\end{eqnarray}
\par
The operator $U(\xi)$
acts through changes of the variables $x^T$ and $p^T$. We denote by
$x^T(\xi)$, $r(\xi)$ and $p^T(\xi)$ the new expressions obtained from
$x^T$, $r$ and $p^T$, respectively, after $U(\xi)$ has acted on them.
We have:
\begin{equation} \label{6e2}
x_{\alpha}^T(\xi)\ =\ U(\xi)x_{\alpha}^T U^{\dagger}(\xi)\ ,\ \ \ \ \ \ \
r(\xi)\ =\ U(\xi)rU^{\dagger}(\xi)\ ,\ \ \ \ \ \ r\ =\ \sqrt{-x^{T2}}\ ,
\end{equation}
from which we deduce the differential equations:
\begin{equation}
\label{6e4}
\frac{\partial x_{\alpha}^T(\xi)}{\partial \xi}\ =\ -f(r(\xi))
x_{\alpha}^T(\xi)\ ,\ \ \ \ \ \ \ \
\frac{\partial r(\xi)}{\partial \xi}\ =\ -f(r(\xi))r(\xi)\ .
\end{equation}
We notice that the variable $x_{\alpha}^T/r$ remains unchanged under
the action of $U(\xi)$, which acts as a local dilatation operator in
$x$-space, and hence it is sufficient to study the variation of $r$.
\par
The solution of Eq. (\ref{6e4}) is:
\begin{equation} \label{6e5}
\int _r^{r(\xi)} \frac{dz}{zf(z)}\ =\ -\xi\ ,
\end{equation}
from which we also deduce:
\begin{equation} \label{6e6}
\frac{\partial r(\xi)}{\partial r}\ =\ \frac{r(\xi)f(r(\xi))}
{rf(r)}\ .
\end{equation}
\par
The action of $U(\xi)$ on the momentum operator $p^T$ is more involved.
The operator $p^T(\xi)$ is no longer parallel to $p^T$ and has
components along $x^T$. For reasons that will become evident below, we
parametrize $p^T(\xi)$ by means of two functions $U_{g\xi}=U_{g\xi}
(r(\xi),r)$ and $T_{g\xi}=T_{g\xi}(r(\xi),r)$ as follows:
\begin{eqnarray} \label{6e7}
p_{\alpha}^T(\xi) &=& U(\xi)p_{\alpha}^TU^{\dagger}(\xi) \nonumber \\
&=& e^{{\displaystyle -2U_{g\xi}}}p_{\alpha}^T + e^{{\displaystyle
-2U_{g\xi}}}\left(e^{{\displaystyle -2T_{g\xi}}}-1\right)\frac{x_{\alpha}
^T}{x^{T2}}x^T.p^T \nonumber \\
& & +i\hbar x_{\alpha}^T \left[ \frac{1}{x^{T2}}e^{{\displaystyle 
-2U_{g\xi}}} \left(e^{{\displaystyle -2T_{g\xi}}}-1\right)
-2(\dot U_{g\xi}+\dot T_{g\xi})e^{{\displaystyle 
-2(U_{g\xi}+T_{g\xi})}}\right]\ .
\end{eqnarray}
[The dot operation is defined in Eq. (\ref{3e15}).] The last term,
proportional to $i\hbar x_{\alpha}^T$, is fixed by the hermiticity
condition. From the definition of $p^T(\xi)$ we obtain the differential
equation:
\begin{eqnarray} \label{6e8}
\frac{\partial p_{\alpha}^T(\xi)}{\partial \xi} &=& f(r(\xi))p_{\alpha}^T
(\xi) + 2\dot f(r(\xi))x_{\alpha}^T(\xi) x^T(\xi).p^T(\xi) \nonumber \\
& &\ \ \ + 5i\hbar x_{\alpha}^T(\xi)\dot f(r(\xi)) + 2i\hbar x^{T2}(\xi)
x_{\alpha}^T(\xi) \ddot f(r(\xi))\ .
\end{eqnarray}
[The dot derivations are with respect to $x^{T2}(\xi)$.] Use in both sides 
of Eq. (\ref{6e8}) of parametrization (\ref{6e7}) leads to differential 
equations concerning the functions $U_{g\xi}$ and $T_{g\xi}$:
\begin{equation}
\label{6e9}
-2\frac{\partial U_{g\xi}}{\partial \xi}\ =\ f(r(\xi))\ ,\ \ \ \ \ \ \ \
-2\frac{\partial T_{g\xi}}{\partial \xi}\ =\ 2x^{T2}(\xi)\dot f(r(\xi))\ .
\end{equation}
[The terms proportional to $i\hbar x_{\alpha}^T$ do not lead to new
conditions.] Taking into account Eq. (\ref{6e4}) and the boundary
condition $p^T(\xi=0)=p^T$, the solutions of Eqs. (\ref{6e9}) are:
\begin{equation} \label{6e12}
U_{g\xi}\ =\ \frac{1}{2} \ln \left(\frac{r(\xi)}{r}\right)\ ,\ \ \ \ \ \ \ 
\ T_{g\xi}\ =\ \frac{1}{2} \ln \left(\frac{f(r(\xi))}{f(r)}\right)\ .
\end{equation}
\par
These also imply the relation:
\begin{equation} \label{6e12a}
4r^2\frac{\partial U_{g\xi}}{\partial r^2}\ =\ e^{{\displaystyle 
2T_{g\xi}}}-1\ .
\end{equation}
\par
In the nonrelativistic limit one has the behaviors
\begin{equation} \label{6e12b}
U_{g\xi}\ =\ \frac{1}{2M}U_{g\xi}^{NR}+O\left(\frac{1}{M^2}\right)\ ,
\ \ \ \ \ 
T_{g\xi}\ =\ \frac{1}{2M}T_{g\xi}^{NR}+O\left(\frac{1}{M^2}\right)\ .
\end{equation}
In this limit, Eq. (\ref{6e12a}) reduces to the relation:
\begin{equation} \label{6e12c}
T_{g\xi}^{NR}\ =\ 2r^2\frac{\partial U_{g\xi}^{NR}}{\partial r^2}\ .
\end{equation}
\par
We next study the action of the operator $U$ on the wave equation 
operator $(\widetilde g_0^{-1}+\widetilde V)$. We first consider the
operator $\widetilde g_0^{-1}$ [Eq. (\ref{3e8})], which is composed of 
the Dirac operators $(\gamma_1.p_1\mp m_1)$ and $(-\gamma_2.p_2\mp m_2)$.
The Dirac operator $(\gamma_1.p_1-m_1)$, say, becomes $(\gamma_1.p_1
-m_1)_{\xi}$, where only the operator $p^T$ has changed, according to 
the transformation law (\ref{6e7}). The operator
$(\gamma_1.p_1-m_1)_{\xi}$ has the same structure as the wave equation
operator (\ref{3e14}), in which $V_1=V_2=V_3=0$ and $U_4=U_{g\xi}$,
$T_4=T_{g\xi}$. [The terms proportional to the matrices
$\sigma_{2\alpha\beta}^{TT}$, present in Eq. (\ref{3e14}), mutually
cancel out when expressions (\ref{6e12}) are used for
$U_{g\xi}$ and $T_{g\xi}$.] A similar conclusion is also obtained with
the Dirac operator $(-\gamma_2.p_2-m_2)_{\xi}$. The two operators
$(\gamma_1.p_1)_{\xi}$ and $(-\gamma_2.p_2)_{\xi}$ (strongly) commute
and hence $(\widetilde g_0^{-1})_{\xi}$ is a well defined operator, in 
which the ordering of the Dirac operators is irrelevant.
\par
The action of the operator $U$ on the potential $\widetilde V$ is obtained
by the replacement in it of $r$ by $r(\xi)$, according to the 
transformations (\ref{6e2}) and (\ref{6e5}):
\begin{equation} \label{6e14p}
\widetilde V_{\xi}\ =\ \widetilde V(r(\xi))\ .
\end{equation}
\par
Examining then the norm of the wave function $\widetilde \Psi_{\xi}$ [Eq.
(\ref{4e11})] (in the kernel of which, after the evaluation of the 
action of $\frac{\partial}{\partial s}$, one uses the equations of motion),
one deduces that the passage to the wave function $\Psi_{\xi}$,
characterized by a norm of the type (\ref{3e12}), is again obtained
with transformations of the type (\ref{3e10})-(\ref{3e11}):
\begin{equation} \label{6e15p}
\widetilde V_{\xi}\ =\ \tanh V_{\xi}\ ,\ \ \ \ \ \ 
\widetilde \Psi_{\xi}\ =\ (\cosh V_{\xi}) \Psi_{\xi}\ .
\end{equation}
\par
The Dirac type wave equations satisfied by $\Psi_{\xi}$ have the same 
structure as Eqs. (\ref{3e14}), in which, however, the potentials
$U_{g\xi}$ and $T_{g\xi}$ have been added up to the existing potentials
$U_4(r(\xi))$ and $T_4(r(\xi))$ of $V_{\xi}$. This feature indicates us
that these wave equations could also have been obtained from the following 
wave equation satisfied by a wave function $\widetilde \Psi_{\xi}'$
defined below:
\begin{equation} \label{6e16p}
(\widetilde g_0^{-1}+\widetilde V_{\xi}')\ \widetilde \Psi_{\xi}'\ =\ 0\ ,
\end{equation}
where $\widetilde V_{\xi}'$ is defined as:
\begin{equation} \label{6e17p}
\widetilde V_{\xi}'\ =\ \tanh V_{\xi}'\ ,
\end{equation}
and $V_{\xi}'$ has the following timelike ($V_{2\xi}'$) and spacelike 
($U_{4\xi}'$, $T_{4\xi}'$) components:
\begin{eqnarray} \label{6e18p}
& & V_{2\xi}'\ =\ V_{2\xi}\ =\ V_2(r(\xi))\ ,\nonumber \\
& & U_{4\xi}'\ =\ U_{4\xi}+U_{g\xi}\ =\ U_4(r(\xi))+U_{g\xi}(r(\xi),r)\ ,
\nonumber \\
& & T_{4\xi}'\ =\ T_{4\xi}+T_{g\xi}\ =\ T_4(r(\xi))+T_{g\xi}(r(\xi),r)\ .
\end{eqnarray}
The wave function $\widetilde \Psi_{\xi}'$ is related to $\Psi_{\xi}$ by 
the transformation:
\begin{equation} \label{6e19p}
\widetilde \Psi_{\xi}'\ =\ (\cosh V_{\xi}') \Psi_{\xi}\ .
\end{equation}
The relationship between $\widetilde \Psi_{\xi}'$ and $\widetilde 
\Psi_{\xi}$ is:
\begin{equation} \label{6e20p}
\widetilde \Psi_{\xi}'\ =\ (\cosh V_{\xi}') (\cosh V_{\xi})^{-1}
\widetilde \Psi_{\xi}\ .
\end{equation}
Therefore, the two wave equation operators $[\widetilde g_0^{-1} +
\widetilde V]_{\xi}$ and $(\widetilde g_0^{-1}+\widetilde V_{\xi}')$
are equivalent:
\begin{equation} \label{6e21p}
[\widetilde g_0^{-1}+\widetilde V]_{\xi}\ \approx\ \widetilde g_0^{-1}
+\widetilde V_{\xi}'\ .
\end{equation}
The advantage of the representation $\widetilde \Psi_{\xi}'$ is that its
wave equation operator has the conventional form (\ref{3e9}) with a
potential $\widetilde V_{\xi}'$ which is local. In this representation,
among the three potentials $V_2$, $U_4$ and $T_4$, only $V_2$ 
has a form invariant transformation law.
If the scalar and pseudoscalar potentials, $V_1$ and $V_3$, were present,
they would transform as $V_2$. (In this case $\widetilde V_{lead}$ 
[Eqs. (\ref{6e1})] should also contain the scalar potential $V_1$.)
\par
The above transformation laws satisfy the group property, as can be 
checked by composing two succesive transformations; hence, they can be
used starting from any gauge.
\par
Let us now return to the explicit expression of the function $f$, Eq. 
(\ref{6e1}). Equation (\ref{6e5}) then yields:
\begin{eqnarray} \label{6e14}
& &\frac{1}{(2x(\xi)-1)}\exp\left(\frac{2}{x(\xi)-1}\right)\ = \frac{1}
{(2x-1)}\exp\left(\frac{2}{x-1}-\xi\right)\ ,\ \ \ \ \ x\ =\ \sqrt{1+\frac
{2\alpha}{P_L r}}\ .\nonumber \\
& &
\end{eqnarray}
\par
It does not seem possible to express $r(\xi)$ in compact form in terms
of $r$ and $\xi$. However, the above equation provides easily the 
asymptotic behaviors of $r(\xi)$:
\subequations
\begin{eqnarray}
\label{6e15a}
& & r(\xi)_{\stackrel{{\displaystyle =}}{r\rightarrow \infty}}
r-\frac{\alpha \xi}{2P_L}+O\left(\frac{1}{r}\right)\ , \\
\label{6e15b}
& & r(\xi)_{\stackrel{{\displaystyle =}}{r\rightarrow 0}}
r e^{{\displaystyle -2\xi}}\left(\frac{1+\sqrt{\frac{P_Lr}{2\alpha}}}
{1+\sqrt{\frac{P_Lr}{2\alpha}}e^{-\xi}}\right)
\exp\big\{4\sqrt{\frac{P_Lr}{2\alpha}}(1-e^{{\displaystyle -\xi}})
\big\} + O(r^2)\ \simeq\ re^{{\displaystyle -2\xi}}\ .\nonumber \\
& &
\end{eqnarray}
\endsubequations
We have plotted, in Fig. 3, the curves $r(\xi)$ (in units of 
$2\alpha/P_L$) for three values of the gauge parameter,
$\xi =-2$ (Yennie gauge), $\xi =0$ (Feynman gauge) and $\xi =1$ 
(Landau gauge). It is observed that the large-distance behavior 
(\ref{6e15a}) is reached very rapidly.
\par
From Eqs. (\ref{6e1}) or Eq. (\ref{6e6}) one also obtains the asymptotic
behaviors of $f(r(\xi))/f(r)$:
\subequations
\begin{eqnarray} 
\label{6e16a}
& & {\frac{f(r(\xi))}{f(r)}} {_{\stackrel{{\displaystyle =}}{r
\rightarrow\infty}} 1}+\frac{\alpha\xi}{2P_L r}+O\left(\frac{1}{r^2}\right)
\ , \\
\label{6e16b}
& & {\frac{f(r(\xi))}{f(r)}} {_{\stackrel{{\displaystyle =}}{r
\rightarrow 0}} 1}+\frac{1}{2}\sqrt{\frac{P_Lr}{2\alpha}}\left[
\frac{1}{1+\sqrt{\frac{P_Lr}{2\alpha}}}-\frac{e^{{\displaystyle
-\xi}}}{1+\sqrt{\frac{P_Lr}{2\alpha}}e^{-\xi}}
+4(1-e^{{\displaystyle -\xi}})\right]+O(r)\ .\nonumber \\
& &
\end{eqnarray}
\endsubequations
\par
Equations (\ref{6e15a})-(\ref{6e16b}), together with Eqs. 
(\ref{6e12}) and (\ref{6e18p}), yield the behaviors of the new potentials
in the corresponding limits. The large-distance expansions (\ref{6e15a})
and (\ref{6e16a}) are particularly relevant in the nonrelativistic limit.
\par
Of particular interest is the short-distance behavior of the effective
potential. According to Eqs. (\ref{6e1})-(\ref{6e2}), the wave function
$\widetilde \Psi_{\xi}(x^T)$ is equal to $\widetilde \Psi(x^T(\xi))$; Eq. 
(\ref{6e15b}) indicates us that $r(\xi)$ behaves like $r$ 
near the origin and therefore the behavior of the wave function 
does not change there under the gauge transformation; this in turn means 
that the dominant short-distance singularity of the effective potential 
has remained the same. These features can also be verified explicitly 
from the wave equations.
\par
To this end, let us consider, for the electromagnetic interaction, 
the Pauli-Schr\"odinger type eigenvalue equation obtained from the wave
equation (\ref{3e14}) (in the c.m. frame) \cite{ms}:
\begin{eqnarray} \label{6e17}
\bigg \{e^{{\displaystyle 4(U_4+T_4)}}\ \big [\ \frac {P^2}{4}
e^{{\displaystyle 4V_2}}&-& \frac{1}{2}(m_1^2+m_2^2) 
+ \frac {(m_1^2-m_2^2)^2}{4P^2}e^{{\displaystyle -4V_2}}\ \big ]
\nonumber \\
&-& {\bf p^2}\ -\ {\bf L^2}\frac {1}{r^2} (e^{{\displaystyle 4T_4}}
-1)\ +\ \cdots \bigg \}\ \phi_3\ =\ 0\ , 
\end{eqnarray}
where $\phi_3$ is a reduced (four-component) wave function and ${\bf L}$
is the orbital angular momentum operator; the dots stand for spin- and
$\hbar$-dependent terms, which are not relevant for the present purpose.
The dominant short-distance singularity is provided by the term 
$e^{{\displaystyle 4(V_2+U_4+T_4)}}$. In the Feynman gauge, [Eqs.
(\ref{3e23})-(\ref{3e24})], it yields the singularity $\alpha^2/r^2$, 
which is attractive and produces at the critical value $\alpha_c
=\frac{1}{2}$ the fall to the center phenomenon. A detailed analysis of this
problem has shown that the theory undergoes at this value of $\alpha$ a 
chiral phase transition \cite{bcs}; such a conclusion has also been reached
from the Bethe-Salpeter equation in the ladder approximation \cite{mf} 
and from lattice theory calculations \cite{br,go}. Considering Eq.
(\ref{6e17}) in the gauge $\xi$, the potentials $V_2$, $U_4$ and $T_4$
become replaced by $V_{2\xi}'$, $U_{4\xi}'$ and $T_{4\xi}'$, respectively.
We find that the modification of the coefficient of the short-distance 
singularity coming from the form invariant part of $V_{2\xi}'+U_{4\xi}'$,
i.e., from $V_{2\xi}+U_{4\xi}$, is cancelled by that coming from the 
form noninvariant part $U_{g\xi}+T_{g\xi}$, and therefore the same 
singularity $\alpha^2/r^2$ emerges again.
\par
One consequence of this result is that the critical coupling constant
$\alpha_c$ has the same value $\frac{1}{2}$ in all gauges. This is
a consistency check of the formalism, since $\alpha$, representing here the
invariant charge, should lead to a gauge invariant 
critical value $\alpha_c$. (In the present approximation, where radiative
corrections are neglected, the physical and bare charges are identical;
in any event, both quantities should be gauge invariant.)
This is in contrast
with the results obtained from the Bethe-Salpeter equation in the ladder
approximation, where the value of $\alpha_c$ is gauge dependent (equal
to $\pi/4$ in the Feynman gauge and to $\pi/3$ in the Landau gauge 
\cite{mf,se}).
\par
We also emphasize the particular role played by $\widetilde V_{lead}$, in
$f$ [Eq. (\ref{6e1})], in the short-distance behavior of the potentials in
the gauge $\xi$. If $\widetilde V_{lead}$ were absent, then the exact
solution of Eq. (\ref{6e5}) would be $r(\xi)=r-\alpha\xi /(2P_L)$
(the same as the asymptotic behavior (\ref{6e15a})),
producing in the Coulomb potential a singularity shifted to the
position $r=\alpha\xi /(2P_L)$. On the other hand, when 
$\widetilde V_{lead}$ is present in
$f$, the function $rf$ (cf. Eq. (\ref{6e5})) vanishes like $r$ when $r$ 
tends to $0$ and, as a result, the singularity of the Coulomb potential 
becomes maintained at the position $r=0$ in the gauge $\xi$.
\par
The results obtained so far with the approximations (\ref{5e11}) are not
qualitatively modified when the dependence on $\xi$ of $\widetilde 
V_{lead,\xi}$ is introduced with an iterative treatment. This is a 
consequence of the asymptotic behaviors (\ref{6e15a})-(\ref{6e15b}) and 
of the fact that the corresponding behaviors of the function $\widetilde 
\Delta (1+\widetilde V_{lead,\xi})^2$ in $f$ [Eq. (\ref{6e1})] are not 
changed.
Similarly, one can also estimate the effect of the operator
$T$ [Eq. (\ref{5e11})]. The integrand in the argument of the exponential
in $T$ is a positive function. The behavior of the integral can be
studied in the asymptotic regions with the aid of relations 
(\ref{6e15a})-(\ref{6e15b}). In the limit $r\rightarrow 0$, the 
argument of the exponential tends to zero, while for $r\rightarrow
\infty$, it tends to the constant value $(\xi_2-\xi_1)\alpha^2
p_{1L}p_{2L}/(2P_L^2)$ and generally remains a smooth function
between these two limits. These properties justify the factorization
approximation made at the level of the gauge transformation operator
$\overline U$ [Eqs. (\ref{5e10}) and (\ref{5e11})].
\par
Let us finally remark that transformations (\ref{6e1}), because of their
spin-independent character, can also be applied in the case of bosonic 
wave equations, where now the operator $\widetilde g_0^{-1}$ is equal 
to $-H_0$, and the distinction between the representations 
$\widetilde \Psi_{\xi}$ and $\widetilde \Psi_{\xi}'$ becomes irrelevant.
\par

\newpage

\section{Summary and concluding remarks}

Using connection with quantum field theory, we established the 
infinitesimal covariant abelian gauge transformation laws
of constraint theory two-particle wave functions and potentials and 
showed weak invariance of the coresponding wave equations. Contrary to
the four-dimensional case and because of the three-dimensional 
projection operation, these transformation laws are interaction
dependent.
\par
Simplifications occur when one sticks to local potentials, 
which result, in each formal order of perturbation theory,
from the infra-red leading effects of multiphoton exchange diagrams.
In this case, the finite gauge transformation can explicitly be represented,
with a suitable approximation and up to a multiplicative factor,
by a momentum dependent unitary operator that acts in $x$-space as
a local dilatation operator. The latter acts on the potentials through
two kinds of modification: a change of the argument $r$ of the
potentials into a function $r(\xi)$ and a functional change of
certain parts of the spacelike components of the electromagnetic potential.
The function $r(\xi)$ is essentially dominated by its large-distance
behavior, in which $r$ is simply shifted by a constant value.
The knowledge of these modifications allows one to reconstruct,
starting from the Feynman gauge, the potentials in other covariant 
gauges. It was shown that the dominant short-distance singularity
of the effective potential of the Pauli-Schr\"odinger type eigenvalue
equation is gauge invariant with a critical value $\alpha_c$ of the
coupling constant equal to $1/2$.
\par
The above results allow the search for optimal gauges when 
incorporating new contributions into the potentials, which might come 
either from QED, or from other interactions. For instance, it is known 
that vacuum polarization affects only the transverse part of the photon 
propagataor. Therefore, the choice of the Landau gauge for the introduction 
of the effective charge seems to be most indicated.
\par
The similarity in structure between QED and perturbative QCD, up to the 
color gauge group matrices and the difference in the effective charges,
allows us to envisage the consideration of many of the previous results 
in problems of quarkonium spectroscopy, where one has also to incorporate at
large distances the effects coming from the confining potential. 
Here also the search for optimal gauges may become useful for subsequent 
applications.
\par

\newpage 

\noindent
{\large {\bf Figures}}
\vspace{1 cm}

\noindent
Fig. 1. Infinitesimal change, under gauge transformations, of two-particle
Green's function. (One-particle radiative corrections are neglected.)
The dashed line, including its vertices, is the gauge propagator 
$ie^2\delta\xi/(k^2)^2$.
\vspace{1 cm}

\noindent
Fig. 2. The three diagrams entering in the evaluation of the
interaction dependent part of the change of the constraint theory wave 
function. The shaded box is the off-mass shell scattering amplitude.
The cross indicates the constraint diagram.
\vspace{1 cm}
 
\noindent
Fig. 3. The curves $r(\xi)$ versus $r$ (in units of $2\alpha/P_L$)
for three values of the gauge parameter, $\xi =-2$ (Yennie gauge),
$\xi =0$ (Feynman gauge) and $\xi =1$ (Landau gauge).

\par

\newpage 

\begin{figure}[h]
\centering 
\epsfysize=9cm 
\epsfbox{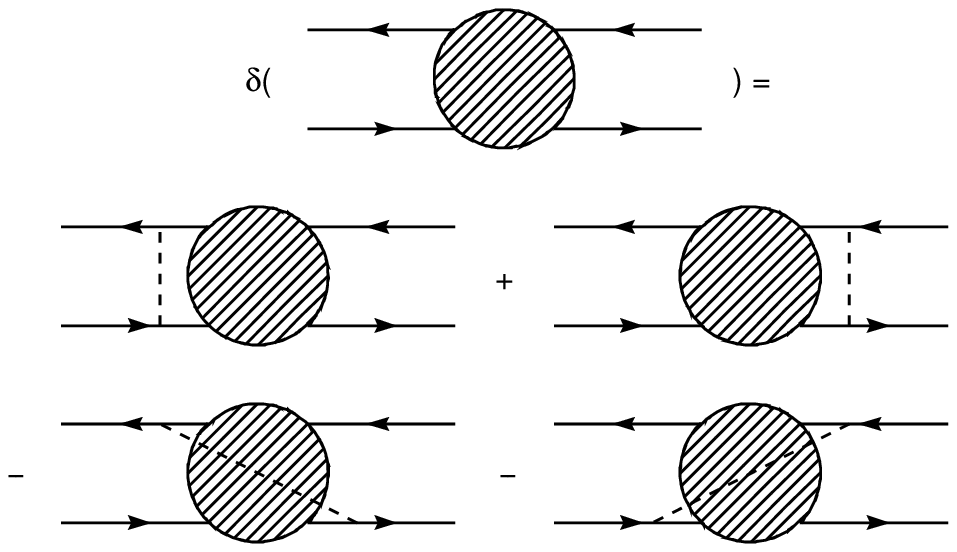}
\label{fig1}
\end{figure}

\begin{center}
Fig. 1
\end{center}

\vspace{1cm}

\begin{figure}[h]
\centering 
\epsfysize=7cm 
\epsfbox{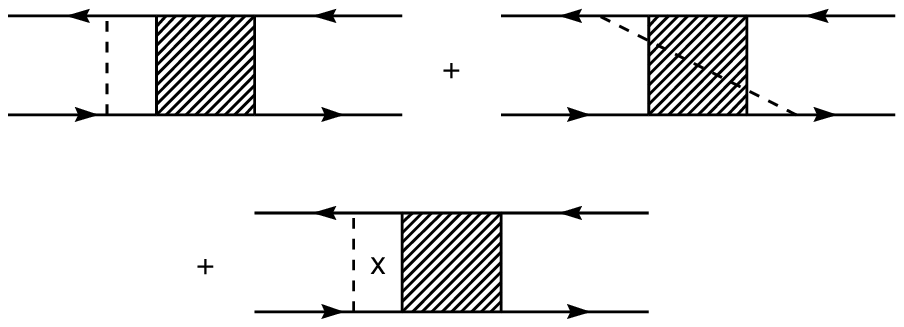}
\label{fig2}
\end{figure}

\begin{center}
Fig. 2
\end{center}

\newpage

\begin{figure}[h]
\begin{center}
\setlength{\unitlength}{0.240900pt}
\ifx\plotpoint\undefined\newsavebox{\plotpoint}\fi
\sbox{\plotpoint}{\rule[-0.200pt]{0.400pt}{0.400pt}}%
\begin{picture}(1800,1350)(0,0)
\font\gnuplot=cmr10 at 10pt
\gnuplot
\sbox{\plotpoint}{\rule[-0.200pt]{0.400pt}{0.400pt}}%
\put(220.0,113.0){\rule[-0.200pt]{365.204pt}{0.400pt}}
\put(220.0,113.0){\rule[-0.200pt]{0.400pt}{292.453pt}}
\put(220.0,113.0){\rule[-0.200pt]{4.818pt}{0.400pt}}
\put(198,113){\makebox(0,0)[r]{0}}
\put(1716.0,113.0){\rule[-0.200pt]{4.818pt}{0.400pt}}
\put(220.0,518.0){\rule[-0.200pt]{4.818pt}{0.400pt}}
\put(198,518){\makebox(0,0)[r]{1}}
\put(1716.0,518.0){\rule[-0.200pt]{4.818pt}{0.400pt}}
\put(220.0,922.0){\rule[-0.200pt]{4.818pt}{0.400pt}}
\put(198,922){\makebox(0,0)[r]{2}}
\put(1716.0,922.0){\rule[-0.200pt]{4.818pt}{0.400pt}}
\put(220.0,1327.0){\rule[-0.200pt]{4.818pt}{0.400pt}}
\put(198,1327){\makebox(0,0)[r]{3}}
\put(1716.0,1327.0){\rule[-0.200pt]{4.818pt}{0.400pt}}
\put(220.0,113.0){\rule[-0.200pt]{0.400pt}{4.818pt}}
\put(220,68){\makebox(0,0){0}}
\put(220.0,1307.0){\rule[-0.200pt]{0.400pt}{4.818pt}}
\put(725.0,113.0){\rule[-0.200pt]{0.400pt}{4.818pt}}
\put(725,68){\makebox(0,0){1}}
\put(725.0,1307.0){\rule[-0.200pt]{0.400pt}{4.818pt}}
\put(1231.0,113.0){\rule[-0.200pt]{0.400pt}{4.818pt}}
\put(1231,68){\makebox(0,0){2}}
\put(1231.0,1307.0){\rule[-0.200pt]{0.400pt}{4.818pt}}
\put(1736.0,113.0){\rule[-0.200pt]{0.400pt}{4.818pt}}
\put(1736,68){\makebox(0,0){3}}
\put(1736.0,1307.0){\rule[-0.200pt]{0.400pt}{4.818pt}}
\put(220.0,113.0){\rule[-0.200pt]{365.204pt}{0.400pt}}
\put(1736.0,113.0){\rule[-0.200pt]{0.400pt}{292.453pt}}
\put(220.0,1327.0){\rule[-0.200pt]{365.204pt}{0.400pt}}
\put(45,720){\makebox(0,0){$r(\xi)$}}
\put(978,23){\makebox(0,0){$r$}}
\put(978,1125){\makebox(0,0)[l]{$\xi =-2$}}
\put(1231,1044){\makebox(0,0)[l]{$\xi =0$}}
\put(1231,720){\makebox(0,0)[l]{$\xi =1$}}
\put(220.0,113.0){\rule[-0.200pt]{0.400pt}{292.453pt}}
\put(220,113){\usebox{\plotpoint}}
\multiput(220.58,113.00)(0.491,3.414){17}{\rule{0.118pt}{2.740pt}}
\multiput(219.17,113.00)(10.000,60.313){2}{\rule{0.400pt}{1.370pt}}
\multiput(230.58,179.00)(0.491,1.486){17}{\rule{0.118pt}{1.260pt}}
\multiput(229.17,179.00)(10.000,26.385){2}{\rule{0.400pt}{0.630pt}}
\multiput(240.58,208.00)(0.491,1.121){17}{\rule{0.118pt}{0.980pt}}
\multiput(239.17,208.00)(10.000,19.966){2}{\rule{0.400pt}{0.490pt}}
\multiput(250.58,230.00)(0.491,0.912){17}{\rule{0.118pt}{0.820pt}}
\multiput(249.17,230.00)(10.000,16.298){2}{\rule{0.400pt}{0.410pt}}
\multiput(260.58,248.00)(0.492,0.732){19}{\rule{0.118pt}{0.682pt}}
\multiput(259.17,248.00)(11.000,14.585){2}{\rule{0.400pt}{0.341pt}}
\multiput(271.58,264.00)(0.491,0.756){17}{\rule{0.118pt}{0.700pt}}
\multiput(270.17,264.00)(10.000,13.547){2}{\rule{0.400pt}{0.350pt}}
\multiput(281.58,279.00)(0.491,0.704){17}{\rule{0.118pt}{0.660pt}}
\multiput(280.17,279.00)(10.000,12.630){2}{\rule{0.400pt}{0.330pt}}
\multiput(291.58,293.00)(0.491,0.652){17}{\rule{0.118pt}{0.620pt}}
\multiput(290.17,293.00)(10.000,11.713){2}{\rule{0.400pt}{0.310pt}}
\multiput(301.58,306.00)(0.491,0.600){17}{\rule{0.118pt}{0.580pt}}
\multiput(300.17,306.00)(10.000,10.796){2}{\rule{0.400pt}{0.290pt}}
\multiput(311.58,318.00)(0.491,0.600){17}{\rule{0.118pt}{0.580pt}}
\multiput(310.17,318.00)(10.000,10.796){2}{\rule{0.400pt}{0.290pt}}
\multiput(321.58,330.00)(0.491,0.600){17}{\rule{0.118pt}{0.580pt}}
\multiput(320.17,330.00)(10.000,10.796){2}{\rule{0.400pt}{0.290pt}}
\multiput(331.58,342.00)(0.491,0.547){17}{\rule{0.118pt}{0.540pt}}
\multiput(330.17,342.00)(10.000,9.879){2}{\rule{0.400pt}{0.270pt}}
\multiput(341.58,353.00)(0.491,0.547){17}{\rule{0.118pt}{0.540pt}}
\multiput(340.17,353.00)(10.000,9.879){2}{\rule{0.400pt}{0.270pt}}
\multiput(351.58,364.00)(0.491,0.547){17}{\rule{0.118pt}{0.540pt}}
\multiput(350.17,364.00)(10.000,9.879){2}{\rule{0.400pt}{0.270pt}}
\multiput(361.00,375.58)(0.547,0.491){17}{\rule{0.540pt}{0.118pt}}
\multiput(361.00,374.17)(9.879,10.000){2}{\rule{0.270pt}{0.400pt}}
\multiput(372.00,385.58)(0.495,0.491){17}{\rule{0.500pt}{0.118pt}}
\multiput(372.00,384.17)(8.962,10.000){2}{\rule{0.250pt}{0.400pt}}
\multiput(382.00,395.58)(0.495,0.491){17}{\rule{0.500pt}{0.118pt}}
\multiput(382.00,394.17)(8.962,10.000){2}{\rule{0.250pt}{0.400pt}}
\multiput(392.00,405.58)(0.495,0.491){17}{\rule{0.500pt}{0.118pt}}
\multiput(392.00,404.17)(8.962,10.000){2}{\rule{0.250pt}{0.400pt}}
\multiput(402.00,415.58)(0.495,0.491){17}{\rule{0.500pt}{0.118pt}}
\multiput(402.00,414.17)(8.962,10.000){2}{\rule{0.250pt}{0.400pt}}
\multiput(412.00,425.58)(0.495,0.491){17}{\rule{0.500pt}{0.118pt}}
\multiput(412.00,424.17)(8.962,10.000){2}{\rule{0.250pt}{0.400pt}}
\multiput(422.00,435.58)(0.495,0.491){17}{\rule{0.500pt}{0.118pt}}
\multiput(422.00,434.17)(8.962,10.000){2}{\rule{0.250pt}{0.400pt}}
\multiput(432.00,445.59)(0.553,0.489){15}{\rule{0.544pt}{0.118pt}}
\multiput(432.00,444.17)(8.870,9.000){2}{\rule{0.272pt}{0.400pt}}
\multiput(442.00,454.58)(0.495,0.491){17}{\rule{0.500pt}{0.118pt}}
\multiput(442.00,453.17)(8.962,10.000){2}{\rule{0.250pt}{0.400pt}}
\multiput(452.00,464.59)(0.611,0.489){15}{\rule{0.589pt}{0.118pt}}
\multiput(452.00,463.17)(9.778,9.000){2}{\rule{0.294pt}{0.400pt}}
\multiput(463.00,473.58)(0.495,0.491){17}{\rule{0.500pt}{0.118pt}}
\multiput(463.00,472.17)(8.962,10.000){2}{\rule{0.250pt}{0.400pt}}
\multiput(473.00,483.59)(0.553,0.489){15}{\rule{0.544pt}{0.118pt}}
\multiput(473.00,482.17)(8.870,9.000){2}{\rule{0.272pt}{0.400pt}}
\multiput(483.00,492.59)(0.553,0.489){15}{\rule{0.544pt}{0.118pt}}
\multiput(483.00,491.17)(8.870,9.000){2}{\rule{0.272pt}{0.400pt}}
\multiput(493.00,501.59)(0.553,0.489){15}{\rule{0.544pt}{0.118pt}}
\multiput(493.00,500.17)(8.870,9.000){2}{\rule{0.272pt}{0.400pt}}
\multiput(503.00,510.59)(0.553,0.489){15}{\rule{0.544pt}{0.118pt}}
\multiput(503.00,509.17)(8.870,9.000){2}{\rule{0.272pt}{0.400pt}}
\multiput(513.00,519.59)(0.553,0.489){15}{\rule{0.544pt}{0.118pt}}
\multiput(513.00,518.17)(8.870,9.000){2}{\rule{0.272pt}{0.400pt}}
\multiput(523.00,528.58)(0.574,0.499){173}{\rule{0.559pt}{0.120pt}}
\multiput(523.00,527.17)(99.840,88.000){2}{\rule{0.280pt}{0.400pt}}
\multiput(624.00,616.58)(0.587,0.499){169}{\rule{0.570pt}{0.120pt}}
\multiput(624.00,615.17)(99.817,86.000){2}{\rule{0.285pt}{0.400pt}}
\multiput(725.00,702.58)(0.601,0.499){165}{\rule{0.581pt}{0.120pt}}
\multiput(725.00,701.17)(99.794,84.000){2}{\rule{0.290pt}{0.400pt}}
\multiput(826.00,786.58)(0.601,0.499){165}{\rule{0.581pt}{0.120pt}}
\multiput(826.00,785.17)(99.794,84.000){2}{\rule{0.290pt}{0.400pt}}
\multiput(927.00,870.58)(0.614,0.499){163}{\rule{0.592pt}{0.120pt}}
\multiput(927.00,869.17)(100.772,83.000){2}{\rule{0.296pt}{0.400pt}}
\multiput(1029.00,953.58)(0.616,0.499){161}{\rule{0.593pt}{0.120pt}}
\multiput(1029.00,952.17)(99.770,82.000){2}{\rule{0.296pt}{0.400pt}}
\multiput(1130.00,1035.58)(0.616,0.499){161}{\rule{0.593pt}{0.120pt}}
\multiput(1130.00,1034.17)(99.770,82.000){2}{\rule{0.296pt}{0.400pt}}
\multiput(1231.00,1117.58)(0.616,0.499){161}{\rule{0.593pt}{0.120pt}}
\multiput(1231.00,1116.17)(99.770,82.000){2}{\rule{0.296pt}{0.400pt}}
\multiput(1332.00,1199.58)(0.616,0.499){161}{\rule{0.593pt}{0.120pt}}
\multiput(1332.00,1198.17)(99.770,82.000){2}{\rule{0.296pt}{0.400pt}}
\multiput(1433.00,1281.58)(0.620,0.498){89}{\rule{0.596pt}{0.120pt}}
\multiput(1433.00,1280.17)(55.764,46.000){2}{\rule{0.298pt}{0.400pt}}
\sbox{\plotpoint}{\rule[-0.400pt]{0.800pt}{0.800pt}}%
\put(220,113){\usebox{\plotpoint}}
\multiput(220.00,114.40)(0.627,0.520){9}{\rule{1.200pt}{0.125pt}}
\multiput(220.00,111.34)(7.509,8.000){2}{\rule{0.600pt}{0.800pt}}
\multiput(230.00,122.40)(0.627,0.520){9}{\rule{1.200pt}{0.125pt}}
\multiput(230.00,119.34)(7.509,8.000){2}{\rule{0.600pt}{0.800pt}}
\multiput(240.00,130.40)(0.627,0.520){9}{\rule{1.200pt}{0.125pt}}
\multiput(240.00,127.34)(7.509,8.000){2}{\rule{0.600pt}{0.800pt}}
\multiput(250.00,138.40)(0.627,0.520){9}{\rule{1.200pt}{0.125pt}}
\multiput(250.00,135.34)(7.509,8.000){2}{\rule{0.600pt}{0.800pt}}
\multiput(260.00,146.40)(0.700,0.520){9}{\rule{1.300pt}{0.125pt}}
\multiput(260.00,143.34)(8.302,8.000){2}{\rule{0.650pt}{0.800pt}}
\multiput(271.00,154.40)(0.548,0.516){11}{\rule{1.089pt}{0.124pt}}
\multiput(271.00,151.34)(7.740,9.000){2}{\rule{0.544pt}{0.800pt}}
\multiput(281.00,163.40)(0.627,0.520){9}{\rule{1.200pt}{0.125pt}}
\multiput(281.00,160.34)(7.509,8.000){2}{\rule{0.600pt}{0.800pt}}
\multiput(291.00,171.40)(0.627,0.520){9}{\rule{1.200pt}{0.125pt}}
\multiput(291.00,168.34)(7.509,8.000){2}{\rule{0.600pt}{0.800pt}}
\multiput(301.00,179.40)(0.627,0.520){9}{\rule{1.200pt}{0.125pt}}
\multiput(301.00,176.34)(7.509,8.000){2}{\rule{0.600pt}{0.800pt}}
\multiput(311.00,187.40)(0.627,0.520){9}{\rule{1.200pt}{0.125pt}}
\multiput(311.00,184.34)(7.509,8.000){2}{\rule{0.600pt}{0.800pt}}
\multiput(321.00,195.40)(0.627,0.520){9}{\rule{1.200pt}{0.125pt}}
\multiput(321.00,192.34)(7.509,8.000){2}{\rule{0.600pt}{0.800pt}}
\multiput(331.00,203.40)(0.627,0.520){9}{\rule{1.200pt}{0.125pt}}
\multiput(331.00,200.34)(7.509,8.000){2}{\rule{0.600pt}{0.800pt}}
\multiput(341.00,211.40)(0.627,0.520){9}{\rule{1.200pt}{0.125pt}}
\multiput(341.00,208.34)(7.509,8.000){2}{\rule{0.600pt}{0.800pt}}
\multiput(351.00,219.40)(0.627,0.520){9}{\rule{1.200pt}{0.125pt}}
\multiput(351.00,216.34)(7.509,8.000){2}{\rule{0.600pt}{0.800pt}}
\multiput(361.00,227.40)(0.700,0.520){9}{\rule{1.300pt}{0.125pt}}
\multiput(361.00,224.34)(8.302,8.000){2}{\rule{0.650pt}{0.800pt}}
\multiput(372.00,235.40)(0.627,0.520){9}{\rule{1.200pt}{0.125pt}}
\multiput(372.00,232.34)(7.509,8.000){2}{\rule{0.600pt}{0.800pt}}
\multiput(382.00,243.40)(0.548,0.516){11}{\rule{1.089pt}{0.124pt}}
\multiput(382.00,240.34)(7.740,9.000){2}{\rule{0.544pt}{0.800pt}}
\multiput(392.00,252.40)(0.627,0.520){9}{\rule{1.200pt}{0.125pt}}
\multiput(392.00,249.34)(7.509,8.000){2}{\rule{0.600pt}{0.800pt}}
\multiput(402.00,260.40)(0.627,0.520){9}{\rule{1.200pt}{0.125pt}}
\multiput(402.00,257.34)(7.509,8.000){2}{\rule{0.600pt}{0.800pt}}
\multiput(412.00,268.40)(0.627,0.520){9}{\rule{1.200pt}{0.125pt}}
\multiput(412.00,265.34)(7.509,8.000){2}{\rule{0.600pt}{0.800pt}}
\multiput(422.00,276.40)(0.627,0.520){9}{\rule{1.200pt}{0.125pt}}
\multiput(422.00,273.34)(7.509,8.000){2}{\rule{0.600pt}{0.800pt}}
\multiput(432.00,284.40)(0.627,0.520){9}{\rule{1.200pt}{0.125pt}}
\multiput(432.00,281.34)(7.509,8.000){2}{\rule{0.600pt}{0.800pt}}
\multiput(442.00,292.40)(0.627,0.520){9}{\rule{1.200pt}{0.125pt}}
\multiput(442.00,289.34)(7.509,8.000){2}{\rule{0.600pt}{0.800pt}}
\multiput(452.00,300.40)(0.700,0.520){9}{\rule{1.300pt}{0.125pt}}
\multiput(452.00,297.34)(8.302,8.000){2}{\rule{0.650pt}{0.800pt}}
\multiput(463.00,308.40)(0.627,0.520){9}{\rule{1.200pt}{0.125pt}}
\multiput(463.00,305.34)(7.509,8.000){2}{\rule{0.600pt}{0.800pt}}
\multiput(473.00,316.40)(0.627,0.520){9}{\rule{1.200pt}{0.125pt}}
\multiput(473.00,313.34)(7.509,8.000){2}{\rule{0.600pt}{0.800pt}}
\multiput(483.00,324.40)(0.548,0.516){11}{\rule{1.089pt}{0.124pt}}
\multiput(483.00,321.34)(7.740,9.000){2}{\rule{0.544pt}{0.800pt}}
\multiput(493.00,333.40)(0.627,0.520){9}{\rule{1.200pt}{0.125pt}}
\multiput(493.00,330.34)(7.509,8.000){2}{\rule{0.600pt}{0.800pt}}
\multiput(503.00,341.40)(0.627,0.520){9}{\rule{1.200pt}{0.125pt}}
\multiput(503.00,338.34)(7.509,8.000){2}{\rule{0.600pt}{0.800pt}}
\multiput(513.00,349.40)(0.627,0.520){9}{\rule{1.200pt}{0.125pt}}
\multiput(513.00,346.34)(7.509,8.000){2}{\rule{0.600pt}{0.800pt}}
\multiput(523.00,357.41)(0.624,0.501){155}{\rule{1.198pt}{0.121pt}}
\multiput(523.00,354.34)(98.514,81.000){2}{\rule{0.599pt}{0.800pt}}
\multiput(624.00,438.41)(0.624,0.501){155}{\rule{1.198pt}{0.121pt}}
\multiput(624.00,435.34)(98.514,81.000){2}{\rule{0.599pt}{0.800pt}}
\multiput(725.00,519.41)(0.624,0.501){155}{\rule{1.198pt}{0.121pt}}
\multiput(725.00,516.34)(98.514,81.000){2}{\rule{0.599pt}{0.800pt}}
\multiput(826.00,600.41)(0.624,0.501){155}{\rule{1.198pt}{0.121pt}}
\multiput(826.00,597.34)(98.514,81.000){2}{\rule{0.599pt}{0.800pt}}
\multiput(927.00,681.41)(0.638,0.501){153}{\rule{1.220pt}{0.121pt}}
\multiput(927.00,678.34)(99.468,80.000){2}{\rule{0.610pt}{0.800pt}}
\multiput(1029.00,761.41)(0.624,0.501){155}{\rule{1.198pt}{0.121pt}}
\multiput(1029.00,758.34)(98.514,81.000){2}{\rule{0.599pt}{0.800pt}}
\multiput(1130.00,842.41)(0.624,0.501){155}{\rule{1.198pt}{0.121pt}}
\multiput(1130.00,839.34)(98.514,81.000){2}{\rule{0.599pt}{0.800pt}}
\multiput(1231.00,923.41)(0.624,0.501){155}{\rule{1.198pt}{0.121pt}}
\multiput(1231.00,920.34)(98.514,81.000){2}{\rule{0.599pt}{0.800pt}}
\multiput(1332.00,1004.41)(0.624,0.501){155}{\rule{1.198pt}{0.121pt}}
\multiput(1332.00,1001.34)(98.514,81.000){2}{\rule{0.599pt}{0.800pt}}
\multiput(1433.00,1085.41)(0.624,0.501){155}{\rule{1.198pt}{0.121pt}}
\multiput(1433.00,1082.34)(98.514,81.000){2}{\rule{0.599pt}{0.800pt}}
\multiput(1534.00,1166.41)(0.624,0.501){155}{\rule{1.198pt}{0.121pt}}
\multiput(1534.00,1163.34)(98.514,81.000){2}{\rule{0.599pt}{0.800pt}}
\multiput(1635.00,1247.41)(0.624,0.501){155}{\rule{1.198pt}{0.121pt}}
\multiput(1635.00,1244.34)(98.514,81.000){2}{\rule{0.599pt}{0.800pt}}
\sbox{\plotpoint}{\rule[-0.200pt]{0.400pt}{0.400pt}}%
\put(220,113){\usebox{\plotpoint}}
\put(220,113.17){\rule{2.100pt}{0.400pt}}
\multiput(220.00,112.17)(5.641,2.000){2}{\rule{1.050pt}{0.400pt}}
\put(230,115.17){\rule{2.100pt}{0.400pt}}
\multiput(230.00,114.17)(5.641,2.000){2}{\rule{1.050pt}{0.400pt}}
\multiput(240.00,117.61)(2.025,0.447){3}{\rule{1.433pt}{0.108pt}}
\multiput(240.00,116.17)(7.025,3.000){2}{\rule{0.717pt}{0.400pt}}
\multiput(250.00,120.61)(2.025,0.447){3}{\rule{1.433pt}{0.108pt}}
\multiput(250.00,119.17)(7.025,3.000){2}{\rule{0.717pt}{0.400pt}}
\multiput(260.00,123.60)(1.505,0.468){5}{\rule{1.200pt}{0.113pt}}
\multiput(260.00,122.17)(8.509,4.000){2}{\rule{0.600pt}{0.400pt}}
\multiput(271.00,127.60)(1.358,0.468){5}{\rule{1.100pt}{0.113pt}}
\multiput(271.00,126.17)(7.717,4.000){2}{\rule{0.550pt}{0.400pt}}
\multiput(281.00,131.60)(1.358,0.468){5}{\rule{1.100pt}{0.113pt}}
\multiput(281.00,130.17)(7.717,4.000){2}{\rule{0.550pt}{0.400pt}}
\multiput(291.00,135.60)(1.358,0.468){5}{\rule{1.100pt}{0.113pt}}
\multiput(291.00,134.17)(7.717,4.000){2}{\rule{0.550pt}{0.400pt}}
\multiput(301.00,139.59)(1.044,0.477){7}{\rule{0.900pt}{0.115pt}}
\multiput(301.00,138.17)(8.132,5.000){2}{\rule{0.450pt}{0.400pt}}
\multiput(311.00,144.59)(1.044,0.477){7}{\rule{0.900pt}{0.115pt}}
\multiput(311.00,143.17)(8.132,5.000){2}{\rule{0.450pt}{0.400pt}}
\multiput(321.00,149.59)(0.852,0.482){9}{\rule{0.767pt}{0.116pt}}
\multiput(321.00,148.17)(8.409,6.000){2}{\rule{0.383pt}{0.400pt}}
\multiput(331.00,155.59)(1.044,0.477){7}{\rule{0.900pt}{0.115pt}}
\multiput(331.00,154.17)(8.132,5.000){2}{\rule{0.450pt}{0.400pt}}
\multiput(341.00,160.59)(0.852,0.482){9}{\rule{0.767pt}{0.116pt}}
\multiput(341.00,159.17)(8.409,6.000){2}{\rule{0.383pt}{0.400pt}}
\multiput(351.00,166.59)(0.852,0.482){9}{\rule{0.767pt}{0.116pt}}
\multiput(351.00,165.17)(8.409,6.000){2}{\rule{0.383pt}{0.400pt}}
\multiput(361.00,172.59)(1.155,0.477){7}{\rule{0.980pt}{0.115pt}}
\multiput(361.00,171.17)(8.966,5.000){2}{\rule{0.490pt}{0.400pt}}
\multiput(372.00,177.59)(0.721,0.485){11}{\rule{0.671pt}{0.117pt}}
\multiput(372.00,176.17)(8.606,7.000){2}{\rule{0.336pt}{0.400pt}}
\multiput(382.00,184.59)(0.852,0.482){9}{\rule{0.767pt}{0.116pt}}
\multiput(382.00,183.17)(8.409,6.000){2}{\rule{0.383pt}{0.400pt}}
\multiput(392.00,190.59)(0.852,0.482){9}{\rule{0.767pt}{0.116pt}}
\multiput(392.00,189.17)(8.409,6.000){2}{\rule{0.383pt}{0.400pt}}
\multiput(402.00,196.59)(0.721,0.485){11}{\rule{0.671pt}{0.117pt}}
\multiput(402.00,195.17)(8.606,7.000){2}{\rule{0.336pt}{0.400pt}}
\multiput(412.00,203.59)(0.852,0.482){9}{\rule{0.767pt}{0.116pt}}
\multiput(412.00,202.17)(8.409,6.000){2}{\rule{0.383pt}{0.400pt}}
\multiput(422.00,209.59)(0.721,0.485){11}{\rule{0.671pt}{0.117pt}}
\multiput(422.00,208.17)(8.606,7.000){2}{\rule{0.336pt}{0.400pt}}
\multiput(432.00,216.59)(0.721,0.485){11}{\rule{0.671pt}{0.117pt}}
\multiput(432.00,215.17)(8.606,7.000){2}{\rule{0.336pt}{0.400pt}}
\multiput(442.00,223.59)(0.852,0.482){9}{\rule{0.767pt}{0.116pt}}
\multiput(442.00,222.17)(8.409,6.000){2}{\rule{0.383pt}{0.400pt}}
\multiput(452.00,229.59)(0.798,0.485){11}{\rule{0.729pt}{0.117pt}}
\multiput(452.00,228.17)(9.488,7.000){2}{\rule{0.364pt}{0.400pt}}
\multiput(463.00,236.59)(0.721,0.485){11}{\rule{0.671pt}{0.117pt}}
\multiput(463.00,235.17)(8.606,7.000){2}{\rule{0.336pt}{0.400pt}}
\multiput(473.00,243.59)(0.721,0.485){11}{\rule{0.671pt}{0.117pt}}
\multiput(473.00,242.17)(8.606,7.000){2}{\rule{0.336pt}{0.400pt}}
\multiput(483.00,250.59)(0.721,0.485){11}{\rule{0.671pt}{0.117pt}}
\multiput(483.00,249.17)(8.606,7.000){2}{\rule{0.336pt}{0.400pt}}
\multiput(493.00,257.59)(0.721,0.485){11}{\rule{0.671pt}{0.117pt}}
\multiput(493.00,256.17)(8.606,7.000){2}{\rule{0.336pt}{0.400pt}}
\multiput(503.00,264.59)(0.626,0.488){13}{\rule{0.600pt}{0.117pt}}
\multiput(503.00,263.17)(8.755,8.000){2}{\rule{0.300pt}{0.400pt}}
\multiput(513.00,272.59)(0.721,0.485){11}{\rule{0.671pt}{0.117pt}}
\multiput(513.00,271.17)(8.606,7.000){2}{\rule{0.336pt}{0.400pt}}
\multiput(523.00,279.58)(0.683,0.499){145}{\rule{0.646pt}{0.120pt}}
\multiput(523.00,278.17)(99.659,74.000){2}{\rule{0.323pt}{0.400pt}}
\multiput(624.00,353.58)(0.656,0.499){151}{\rule{0.625pt}{0.120pt}}
\multiput(624.00,352.17)(99.703,77.000){2}{\rule{0.312pt}{0.400pt}}
\multiput(725.00,430.58)(0.648,0.499){153}{\rule{0.618pt}{0.120pt}}
\multiput(725.00,429.17)(99.717,78.000){2}{\rule{0.309pt}{0.400pt}}
\multiput(826.00,508.58)(0.639,0.499){155}{\rule{0.611pt}{0.120pt}}
\multiput(826.00,507.17)(99.731,79.000){2}{\rule{0.306pt}{0.400pt}}
\multiput(927.00,587.58)(0.646,0.499){155}{\rule{0.616pt}{0.120pt}}
\multiput(927.00,586.17)(100.721,79.000){2}{\rule{0.308pt}{0.400pt}}
\multiput(1029.00,666.58)(0.631,0.499){157}{\rule{0.605pt}{0.120pt}}
\multiput(1029.00,665.17)(99.744,80.000){2}{\rule{0.303pt}{0.400pt}}
\multiput(1130.00,746.58)(0.631,0.499){157}{\rule{0.605pt}{0.120pt}}
\multiput(1130.00,745.17)(99.744,80.000){2}{\rule{0.303pt}{0.400pt}}
\multiput(1231.00,826.58)(0.631,0.499){157}{\rule{0.605pt}{0.120pt}}
\multiput(1231.00,825.17)(99.744,80.000){2}{\rule{0.303pt}{0.400pt}}
\multiput(1332.00,906.58)(0.624,0.499){159}{\rule{0.599pt}{0.120pt}}
\multiput(1332.00,905.17)(99.757,81.000){2}{\rule{0.299pt}{0.400pt}}
\multiput(1433.00,987.58)(0.631,0.499){157}{\rule{0.605pt}{0.120pt}}
\multiput(1433.00,986.17)(99.744,80.000){2}{\rule{0.303pt}{0.400pt}}
\multiput(1534.00,1067.58)(0.624,0.499){159}{\rule{0.599pt}{0.120pt}}
\multiput(1534.00,1066.17)(99.757,81.000){2}{\rule{0.299pt}{0.400pt}}
\multiput(1635.00,1148.58)(0.631,0.499){157}{\rule{0.605pt}{0.120pt}}
\multiput(1635.00,1147.17)(99.744,80.000){2}{\rule{0.303pt}{0.400pt}}
\end{picture}
\label{fig3}
\end{center}
\end{figure}

\begin{center}
Fig. 3
\end{center}

\end{document}